\documentclass[twoside,twocolumn,english,showpacs,preprintnumbers,amsmath,amssymb]{revtex4}
\usepackage[OT1]{fontenc}
\usepackage[latin1]{inputenc}
\usepackage[dvips]{color}
\usepackage[dvips]{graphicx}
\usepackage{amssymb}

\makeatletter

\providecommand{\tabularnewline}{\\}

\usepackage{amsfonts}

\usepackage{floatflt}

\usepackage{subfigure}

\usepackage{babel}
\makeatother
\begin{document}

\title{Simulating weak lensing on CMB maps}

\author{S. Basak, S. Prunet and K. Benabed}

\affiliation{Institut d'Astrophysique de Paris, CNRS, UMR 7095, 98bis Bld. Arago,
75014 Paris, France. }

\begin{abstract}
We present a fast, arbitrarily accurate method to simulate the effect
of gravitational lensing of the Cosmic Microwave Background anisotropies
and polarization fields by large scale structures. We demonstrate the efficiency and accuracy of the method and exhibit their dependence on the algorithm parameters.


\end{abstract}
\maketitle

\section{Introduction}

Weak lensing effects on the Cosmic Microwave Background (CMB) temperature and polarization anisotropies
has been proposed as a probe of the total matter distribution in Large
Scale Structures (LSS) between us and the surface of last scattering \cite{blanchard1987,seljak1998,zaldarriaga1998,hu1999a,hu1999b,zaldarriaga1999,guzik2000,waerbeke2000,benabed2001,hu2000,kesden2002,hirata2003b,hirata2003a,kesden2003,hirata2004,lewis2005,challinor2005,lewis2006,smith2007}.
Although sensitive to the cumulative distribution of matter, it is
quite complementary to the other probes of the matter distribution
of the LSS. Indeed, it does not suffer from bias effects (as e.g.
galaxy redshift surveys, Lyman-$\alpha$ forest), or from possible
mis-determination of the redshift sources (cosmic shear measurements
on galaxies). In addition, due to the high redshift of the source
(last scattering surface) and the lensing efficiency function, weak
lensing of CMB
anisotropies is mostly sensitive to large scale structures
which are still (mainly) in the linear regime, which makes it a very
useful tool for cosmology, in particular to constrain the properties
of neutrinos \cite{perotto2006,lesgourgues2006}.

Unlike shear measurements on galaxies, where the (reduced) shear field
is directly sampled via measurement of galaxy ellipticities, measuring
weak lensing effects on the CMB is complicated by the fact that the
source itself can only be modeled by a stochastic realization of a field. 
However, theoretical
arguments lead us to think that CMB anisotropies are highly Gaussian
\cite{guth1981,linde1982,albrecht1982}, which has been confirmed
on the data, at large scales,  using different non-Gaussianity estimators (pdf, bispectrum,
wavelet skewness and kurtosis, Minkowski functionals, etc.). These
properties of the CMB anisotropies can be used to disentangle to some
extent the stochastic properties of the (unlensed) CMB anisotropies
from the stochastic properties of the lens (i.e. the LSS) as the lensing
effect induces small specific non-Gaussian features in the CMB maps: it
locally correlates the anisotropies with their gradient \cite{seljak1998,cooray2001a,cooray2001b,cooray2001c,cooray2002}
which in turn has lead to the development of specific estimators of
the lensing potential field and its power spectrum \cite{hu2002,hirata2003b,hirata2003a,okamoto2003}.

Recently, weak lensing of the CMB anisotropies by LSS has been
measured on WMAP data \cite{smith2007} by cross-correlation with
a high-redshift radio galaxy catalog. Although marginally detectable
in WMAP data due to its noise level, CMB lensing should be measured
with high signal to noise by Planck with temperature anisotropies \cite{hu2002,hu2004}
without needing to rely on an external data set. However, in order
to carry out such a measurement on realistic CMB data, the impact
of instrumental (anisotropic beams, missing data, correlated noise)
and astrophysical (Galaxy contamination, point sources, etc.) systematic
effects on the CMB lensing estimators has to be studied with great
care.

The power spectra (temperature, and polarization) can be computed, using
simple Taylor expansion at large scales \cite{hu2000}, or a more clever
resummation scheme at smaller scales where the displacement field
amplitude is comparable to the wavelength of the anisotropies
\cite{challinor2005}.  For smaller scales, or to investigate the
different systematics described above, the development of fast and
accurate methods to simulate the lensed CMB maps are needed.

This simulation is two-fold. On one hand, an accurate simulation of the
large scale structure induced lensing deflection field is needed. On the
other hand, one needs a method to apply this deflection field to an
unlensed, simulated, CMB. We will not consider the first part of this
program. Indeed, approximating the lensing effect with a single lens
plane in the so-called Born approximation \cite{hu2000} has been shown
to be an excellent approximation, both for temperature and polarization
anisotropies. In this case, the simulation of lensed CMB maps reduces to
an accurate resampling of the unlensed anisotropies at displaced
positions. To solve this last problem, several technical solutions have
been implemented. In the publicly distributed Lenspix code
\cite{lenspix}, different possibilities are available, namely:

\begin{itemize}
\item brute-force resampling by direct resummation of spherical harmonics
at displaced positions (slow, but very accurate, this option should
be considered as the {}``benchmark'' for all other resampling methods) 
\item resampling on locally Cartesian grids with subsequent polynomial interpolation 
\end{itemize}
For the last option, an interesting speed-up has been proposed by
Hirata \cite{hirata2004,das_bode2008} by noting that a band-limited signal in
spherical harmonics can be recast as a band-limited signal in regular
Fourier modes on a $(\theta,\varphi)$, thus allowing a fast resampling
of the signal on a Cartesian $(\theta,\varphi)$ grid using 2D FFTs.

In this paper, we investigate a variation on Hirata's idea
\cite{hirata2004,das_bode2008}, where the oversampling plus polynomial
interpolation is replaced by an approximate (but arbitrarily accurate)
Fast Fourier Transform (FFT) resampling on irregularly spaced grid
points \cite{nfft}.

 The rest of this paper is organized as follows: In section \ref{neq_fft}
 we briefly describe the resampling technique (hereafter NFFT). 
 This is followed by a brief
 description of the weak lensing of primary CMB fields in section
 \ref{cmb_rad}. We also describe how remapping of CMB fields on the
 surface of the unit sphere can be recast into remapping of the latter on
 the surface of a 2-d torus.  Section \ref{simulation} describes the
 details of the simulation procedure for lensed CMB fields using NFFT on the
 surface of a 2-d torus. Finally we summarize our results in section
 \ref{summary}.

\section{Non-equispaced fast Fourier transform(NFFT)}
\label{neq_fft} The fast Fourier transform for non-equispaced grid
points (NFFT) is a generalization of FFT
\cite{kunis_potts2008,fourmont_2003}. The essential idea is that of
approximating the reproducing kernel of the standard FFT \cite{fftw}
using a window function of specific properties. Suppose we know a
function $f$ through $N$ evaluation $f_k$ in the frequency
domain. According to NFFT, Fourier transform of that function evaluated
at $M$ non-equispaced grid points in spatial domain can be written as,

\begin{eqnarray}
{\hat f}(x_{j}) = \frac{1}{\sqrt{2\pi}}\sum_{m\in\mathbb{Z}} {\hat
 \phi}(\sigma x_{j}-m)\hspace{1.0in}\nonumber\\
\times\sum^{N/2-1}_{k=-N/2}\exp\left[-\frac{2\pi im\,k}{\sigma\,N}\right]
 \frac{f_{k}}{\phi(2\pi\,k/\sigma\,N)}\\
\nonumber\\
j=1,2,3,\ldots,M\nonumber
\label{nfft1d}
\end{eqnarray}

Here the window function $\phi(\xi)$ has compact support
$\left[-\alpha,\alpha\right]$ and its Fourier transform ${\hat \phi(x)}$
assumes small values outside some interval $\left[-K,K\right]$. $\sigma$
is the over-sampling factor and it is required to avoid the aliasing
error. A convenient choice for $\sigma$ is 2, however $\sigma = 3/2$ is
sufficient to get good accuracy.  $\alpha$ has to be chosen slightly
smaller than $\pi\left(2-2/\sigma\right)$. Since the evaluation of the
summation over $k$ requires an equispaced FFT of length $\sigma\,N$,
$\phi(\xi)$ has to be well localized in k-space in order to avoid the
aliasing error with minimal computational cost. On the other hand, the
summation over $m$ can be evaluated with minimum truncation error if
the window function is well localized in the spatial domain. Hence, the
efficient evaluation of ${\hat f}(x)$ on irregularly spaced grid points
requires a window function that is well localized in both space and
frequency domain. It has computational complexity ${\cal O}\left(\sigma
N\log\,N + K\,M\right)$ where $K$ is the number of terms considered in
the spatial approximation, $M$ is the number of real space samples,
and $N$ the number of Fourier modes. Among a number of window functions (Gaussian,
B-spline, Sinc-power, Kaiser-Bessel), Kaiser-Bessel turns out to be the
best. It has been shown that for a fixed oversampling factor $\sigma >
1$, the approximation error decays exponentially with $K$
\cite{kunis_potts2008,fourmont_2003}.

\section{Weak lensing of CMB}

\label{cmb_rad} The CMB radiation field is completely characterized
by its temperature anisotropy, $T(\theta,\varphi)$ , and polarization,
$P(\theta,\varphi)$, in each direction on the sky. Since temperature
anisotropy is a spin-0 field on the sphere, it can be
conveniently expanded in spin-0 spherical harmonics,

\begin{eqnarray}
T(\theta,\varphi) & = & \sum_{l=0}^{l_{max}}\sum_{m=-l}^{l}T_{lm}\,
 Y_{lm}(\theta,\varphi)
\label{temp_sphere}
\end{eqnarray}

The polarization field can be described by the
Stokes parameters, $Q(\theta,\varphi)$ and $U(\theta,\varphi)$,
with respect to a particular choice of coordinate system on the
sky. One can conveniently combine the Stokes parameters into a single
complex quantity representing the polarization, $P(\theta,\varphi)=\left(Q+iU\right)(\theta,\varphi)$.
Due to its transformation properties under rotations, the polarization
$P$ is a spin-2 field
on the sphere. One may thus expand $P(\theta,\varphi)$
in terms of spin-2 spherical harmonics, ${}_{2}Y_{lm}(\theta,\varphi)$
\cite{zaldarriaga1997,newman1996,goldberg1967}, as

\begin{eqnarray}
P(\theta,\varphi) & = & (Q+i\, U)(\theta,\varphi)\nonumber \\
 & = &
  \sum_{l=0}^{l_{max}}\sum_{m=-l}^{l}{}_{2}P_{lm}\,\,{}_{2}Y_{lm}(\theta,\varphi)
\label{pol_sphere}
\end{eqnarray}
In the above equation, ${}_{2}P_{lm}=-(E_{lm}+i\,B_{lm})$, where
$E_{lm}$ and $B_{lm}$ are the electric and magnetic modes of the
polarization field in harmonic space.

Weak lensing induces a deflection field $\vec{d}(\theta,\varphi)$, i.e. a
mapping between the direction of a given light ray on the last scattering
surface and the direction in which we observe it. Since the deflection field
is a vector field on the sphere, it can be decomposed in
terms of gradient-free and curl-free components in the most general form
as, \begin{eqnarray} d_{a}(\theta,\varphi)  = 
\nabla_{a}\Phi(\theta,\varphi)+\epsilon_{a}^{\,\,
b}\,\nabla_{b}\chi(\theta,\varphi)\end{eqnarray} 
\begin{eqnarray}
\hspace{2.5in}a,b\in(\theta,\phi)\nonumber
\end{eqnarray} 
where
$\Phi(\theta,\varphi)$ and $\chi(\theta,\varphi)$ are two scalar fields
on the sphere. $\epsilon_{a\,b}$ is the covariant
antisymmetric tensor of rank $2$ on the unit sphere. In terms
of null basis vectors $(m,\bar{m})$ which define a diad on the unit sphere, $\epsilon_{a\,b}$
can be expressed as,
\begin{eqnarray}
\epsilon_{a\,b}=i(m_{a}\bar{m}_{b}-\bar{m}_{a}m_{b})
\end{eqnarray}

The gradient-free component can be ignored as it is negligible in most cases \cite{cooray2002}
and is exactly zero in the Born approximation that we use here, as this term
can only arise when taking into account the lens-lens couplings.
In the Born approximation, the lensing deflection is calculated on the
unlensed line of sight so the lensed map is a local function of the
deflection vector,
$d_{a}(\theta,\varphi)=\nabla_{a}\,\Phi(\theta,\varphi)$, where
$\Phi(\theta,\varphi)$ is the lensing potential. This projected
$\Phi(\theta,\varphi)$ potential is related to the 3-d the gravitational
potential $\Psi(D,{\vec D}{(D, \theta, \varphi}))$ as,

\begin{eqnarray}
\Phi(\theta,\varphi) = -2\int^{D_{s}}_{0}dD
 \frac{D_{A}\left(D_{s}-D\right)}{D_{A}\left(D\right)
\,D_{A}\left(D_{s}\right)}\,\Psi(D,{\vec D}{(D, \theta,
\varphi}))\hspace{0.1in}
\end{eqnarray}

where $D$ is the comoving coordinate distance along the line of sight
and $D_{A}$ is the comoving angular diameter distance associated with
D. $D_{s}$ is the coordinate distance to the last scattering surface.

Similarly to CMB temperature anisotropy, the lensing potential transforms like
a spin-zero field on the sphere. Hence it may also be
expanded in spin-0 spherical harmonics.

\begin{eqnarray}
\Phi(\theta,\varphi) & = &
 \sum_{l=0}^{l_{max}}\sum_{m=-l}^{l}\Phi_{lm}\, Y_{lm}(\theta,\varphi)
\label{pot_sphere}
\end{eqnarray}

Since the deflection field $\vec{d}(\theta,\varphi)$ is a vector field
on the sphere, it can be expanded in spin-1 spherical
harmonics,

\begin{eqnarray}
d_{a}(\theta,\varphi) & = & \nabla_{a}\,\Phi(\theta,\varphi)\nonumber \\
 & = & \sum_{l=0}^{l_{max}}\sum_{m=-l}^{l}\Phi_{lm}\,\sqrt{\frac{l(l+1)}{2}}\,\nonumber \\
 & \times & \left[{}_{(-1)}Y_{lm}(\theta,\varphi)\,
	     m_{a}-{}_{1}Y_{lm}(\theta,\varphi)\,\overline{m}_{a}\right]
\label{deflec_sphere}
\end{eqnarray}

NFFT in $2$-dimensions works on $2$-d torus, we have thus rewritten
equations(\ref{temp_sphere}-\ref{deflec_sphere}) into a form (Appendix
\ref{cmb_torus}) that is suitable to simulate unlensed CMB maps at
irregularly spaced grid points using NFFT. This is possible because a
band-limited function on a unit sphere can be rewritten as a
band-limited function on a $2$-d torus. In order to do this, we have
exploited the relation of spin-weighted spherical harmonics to Wigner
rotation matrices(\ref{harm_slm}) and the factorization of Wigner rotation
matrices into two separate rotations(\ref{wig_decomp}).

Using the identities of the spherical triangle, lensed temperature
anisotropies and polarization in a particular direction
$(\theta,\varphi)$ are given by unlensed temperature anisotropies and
polarization in another direction at the last scattering
surface.\begin{eqnarray} \tilde{T}(\theta,\varphi) & = &
T(\theta',\varphi')\label{lensed_map-a}\\ \tilde{P}(\theta,\varphi) & = &
\exp\left[-2i(\gamma-\alpha)\right]\,\,
P(\theta',\varphi')\label{lensed_map-b}\end{eqnarray} The angular
coordinates corresponding to the modified direction of the photon path
$(\theta',\varphi')$ due to lensing are determined by the deflection field
$\vec{d}(\theta,\varphi)$,\begin{eqnarray} \cos\theta' & = & \cos
d\,\cos\theta-\sin d\,\sin\theta\,\cos\alpha \label{sph_triangle-a}\\
\sin(\varphi'-\varphi) & = &
\frac{\sin\alpha\,\sin(d)}{\sin\theta'}\label{sph_triangle-b}\end{eqnarray}
The extra factor $\exp\left[-2i(\gamma-\alpha)\right]$, that appears in
case of polarization {[\ref{lensed_map-b}], is there to rotate the basis vectors
$\left(\hat{e}_{\theta'},\hat{e}_{\phi'}\right)$ at $(\theta',\phi')$ to
match them with the basis vectors
$\left(\hat{e}_{\theta},\hat{e}_{\phi}\right)$ at
$(\theta,\varphi)$.\begin{eqnarray} A=\tan(\gamma) & = &
\frac{d_{\phi}}{d\,\sin d\,\cot\theta+d_{\theta}\, cos\, d}\label{rot_equ-a} \\
\cos\left[2(\alpha-\gamma)\right] & = & \frac{2\left(d_{\theta}+A\,
d_{\phi}\right)^{2}}{d^{2}(1+A^{2})}-1\label{rot_equ-b}  \\
\sin\left[2(\alpha-\gamma)\right] & = & \frac{2\left(d_{\theta}+A\,
d_{\phi}\right)\left(d_{\phi}-A\,
d_{\theta}\right)}{d^{2}(1+A^{2})}\label{rot_equ-c}\end{eqnarray}

The Euler angles $\alpha,\,\beta$ and $\gamma$ are defined as,
\begin{eqnarray} D_{s\, s'}^{l}(\alpha,\beta,-\gamma) =
\sum_{m=-l}^{l}\,\frac{4\pi}{2\,
l+1}\,{}_{s}Y_{lm}^{*}(\theta,\varphi)\,\,{}_{{s}'}Y_{lm}(\theta',\varphi')\hspace{0.2in}\end{eqnarray}
$\beta\,(0\le\beta\le\pi)$ determines the angle between the directions
$(\theta,\varphi)$ and
$(\theta',\varphi')$. $\alpha\,(0\le\alpha\le2\pi)$ is the angle
required to rotate the basis vector
$\hat{\theta}\,\equiv\,(\theta,\varphi)$ in a right-handed sense about
$\hat{n}$ onto the tangent (at $\hat{n}$ ) to the geodesic connecting
$\hat{n}$ and $\hat{n}'$; $\gamma\,(0\le\gamma\le2\pi)$ is defined in
the same manner as $\alpha$ but at $\hat{n}'$.

To compute lensed CMB fields at a particular position on the sphere 
it is enough to compute the unlensed CMB at some other position on
the sphere determined by the identities of the spherical
triangle. The most popular pixelization scheme that is used in
CMB analysis is the
HEALPix\footnote{\texttt{http://healpix.jpl.nasa.gov}} 
 pixelization~\citep{healpix} which is an irregular grid on
the surface of the unit sphere in $(\theta,\phi)$ coordinates. Since
gravitational lensing remaps the CMB signal, the modified angular
coordinates due to lensing will not, in general, correspond to any other pixel
center of the HEALPix grid, even if the unlensed CMB is defined over
HEALPix grid points.  Hence, in order to compute lensed CMB field on
HEALPix grid points, we should be able to resample the unlensed CMB at arbitrary
positions on the sphere.  Since remapping on a sphere can be recast into
remapping on a 2-d torus (Appendix \ref{spin_function}), we have used
NFFT to compute lensed CMB anisotropies at HEALPix grid points.

\section{Simulation of lensed CMB map}

\label{simulation}

\subsection{How to simulate a lensed map\label{sub:Algorithm}}

We have seen in the last section that, in the Born approximation,
gravitational lensing of the CMB anisotropies results in a simple
resampling of the unlensed anisotropies, with an extra rotation in
the case of polarization lensing. Let us summarize here the main steps
of the simulation procedure of lensed CMB maps:

\begin{itemize}
\item Generate a realization of the (unlensed) CMB harmonic coefficients
(both temperature and polarization) from their (unlensed) power spectra
\item Generate in the same way the harmonic coefficients of the lensing
potential, or alternatively extract them from an N-body simulation
\item Transform the harmonic coefficients of the unlensed CMB fields into
their 2-d torus Fourier counterparts using equations
      (\ref{temp_harm_torus} and \ref{pol_harm_torus}).
Also get the Fourier coefficients of the displacement field from the
harmonic coefficients of the lensing potential using equation (\ref{field_harm_torus}).
\item Sample the displacement field at HEALPix centers (using equation
  \ref{defl_torus}
and NFFT), apply this displacement field to HEALPix pixel centers
to get displaced positions on the sphere (using equations
      \ref{sph_triangle-a} \& \ref{sph_triangle-b}).
Also compute the extra rotation that will be needed for the polarized
fields (using equations \ref{rot_equ-a},\ref{rot_equ-b} \& \ref{rot_equ-c}).
\item Resample the temperature and polarization fields at the displaced
positions using equations (\ref{temp_torus} and \ref{pol_torus}) and
      NFFT, apply the extra
rotation to the polarized fields. This gives us the simulated lensed
CMB fields, sampled at HEALPix pixel centers.
\end{itemize}

\subsection{Validation of the method on a known case: unlensed maps}

In order to test the part of the algorithm that goes from harmonic
coefficients of temperature or polarization fields to arbitrary real
space sampled positions, via 2-d torus Fourier modes and NFFT transform,
we test the method on unlensed temperature or polarization fields,
sampled at HEALPix centers. Indeed, this is a valid test of the method
as HEALPix pixel centers are irregularly distributed in $(\theta,\varphi)$
coordinates. In addition, we can directly compare the output of the
method to a direct resummation of the spherical harmonics decomposition
of the fields at HEALPix centers by using the fast spherical harmonics
transforms of the HEALPix package, which will serve as a reference.%
\begin{figure}[h]
\begin{centering}
\includegraphics[scale=0.25,angle=90]{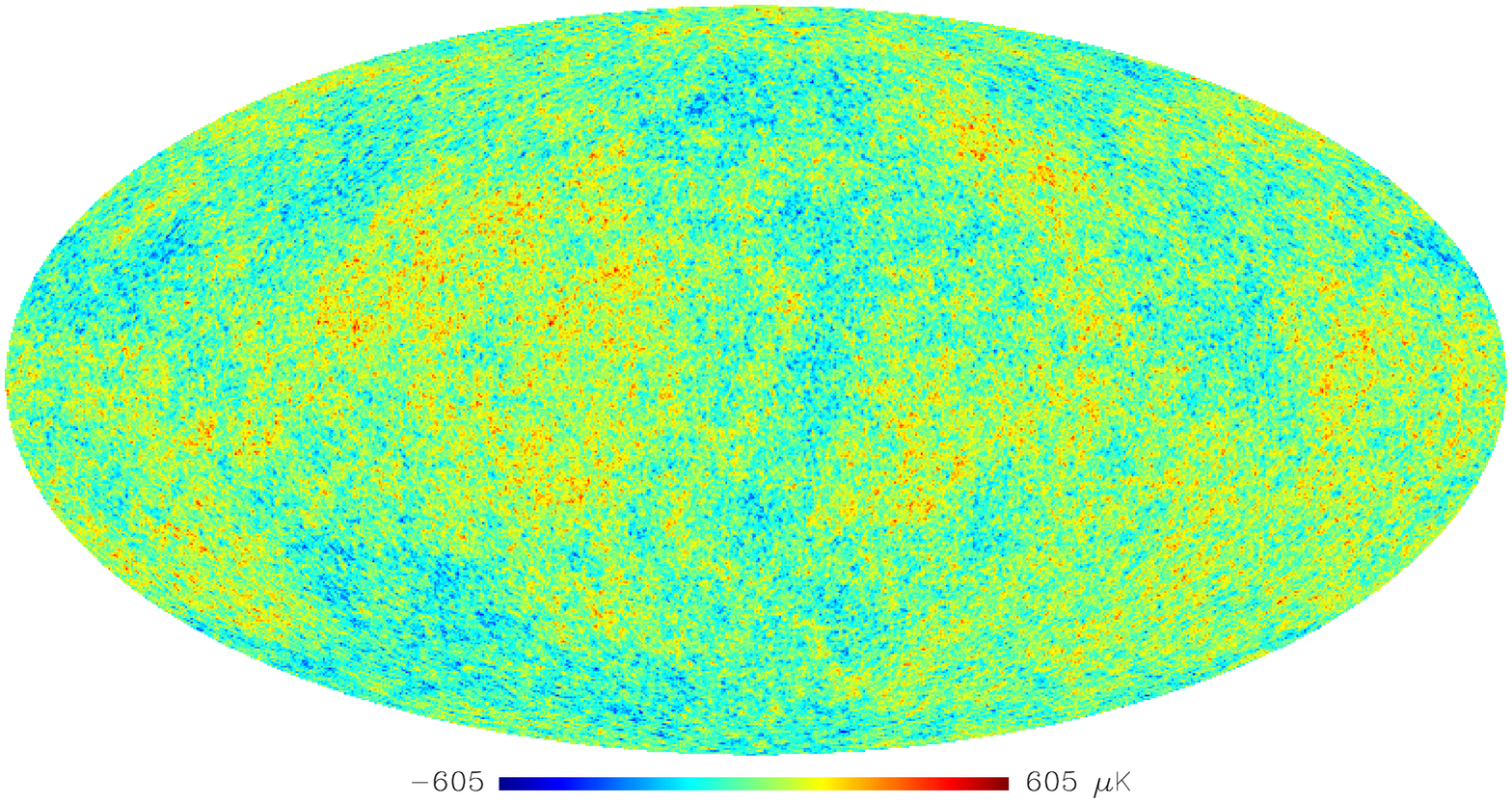}
\par\end{centering}

\begin{centering}
\includegraphics[scale=0.25,angle=90]{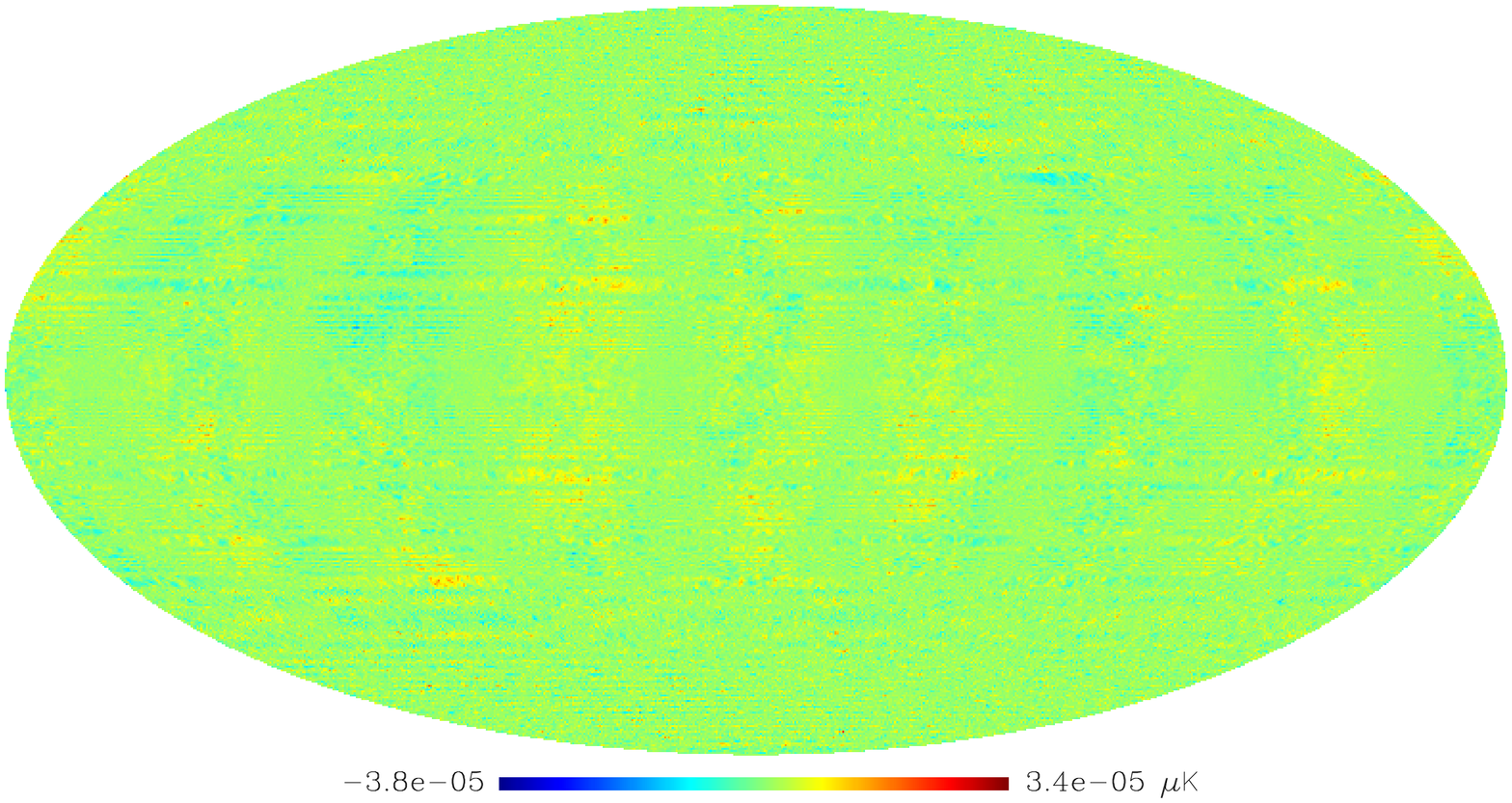}
\par\end{centering}

\caption{(a) A realization of an unlensed CMB temperature anisotropies map
(nside=1024) that we have obtained using NFFT [oversampling factor
($\sigma$)=2, Convolution length $(K)=4$]. (b) Difference of unlensed
CMB maps (nside=1024) that we have obtained using NFFT and HEALPix from
the same harmonic coefficients.}

\label{unlensed_CMB_map_nside1024} 
\end{figure}

In Figure~\ref{unlensed_CMB_map_nside1024}, we show an (unlensed)
realization of the CMB temperature anisotropies obtained using our
method, as well as a map of the difference between our method and
the HEALPix reference map. Note the difference in the color scales.
In order to quantify more precisely the accuracy of our method, we
have computed two kinds of error statistics: \begin{eqnarray*}
E_{\infty}^{X} & = &
 \frac{max_{j}\left|X_{NFFT}(\theta_{j},\phi_{j})-X_{HEALPix}(\theta_{j},\phi_{j})\right|}{max_{j}\left|X_{HEALPix}(\theta_{j},\phi_{j})\right|}\\
\\
E_{2}^{X} & = & \sqrt{\frac{\sum_{j=1}^{Npix}\left|X_{NFFT}(\theta_{j},\phi_{j})-X_{HEALPix}(\theta_{j},\phi_{j})\right|^{2}}{\sum_{j=1}^{Npix}\left|X_{HEALPix}(\theta_{j},\phi_{j})\right|^{2}}}\end{eqnarray*}
where $X$ stands for $T$, $Q$, $U$, $d_{\theta}$ and $d_{\phi}$.

$E_{\infty}^{X}$ is the maximum (relative) error for field X, while
$E_{2}^{X}$ is the relative root mean square
error. Table~\ref{Error_norms} gives the value of these statistics for
unlensed CMB temperature only. Values of these error norms for the
displacement field and the unlensed CMB polarization fields are of the same
order of magnitude.

\begin{table}[h]

\caption{Variation of typical order of magnitude of error norms with the
 convolution length $(K)$ for an unlensed CMB map simulated using
 NFFT. $E_{2}$ and $E_{\infty}$ are the quadratic norm and infinite norm
 respectively}
\begin{centering}
\begin{tabular}{|c|c|c|c|c|c|}
\hline 
Oversampling&
Convolution&
nside&
$l_{max}$&
Maximum&
R. M. S
\tabularnewline
factor&
length&
&
&
error&
error
\tabularnewline
$\left(\sigma\right)$&
$\left(K\right)$&
&
&
$\left(E_{\infty}\right)$&
$\left(E_{2}\right)$
\tabularnewline
\hline 
2&
4&
1024&
2048&
$\sim 10^{-8}$&
$\sim 10^{-8}$
\tabularnewline
\hline 
2&
6&
1024&
2048&
$\sim 10^{-11}$&
$\sim 10^{-12}$
\tabularnewline
\hline
2&
8&
1024&
2048&
$\sim 10^{-11}$&
$\sim 10^{-13}$
\tabularnewline
\hline 
\end{tabular}
\label{Error_norms} 
\par\end{centering}
\end{table}

To achieve this accuracy, we have used the Kaiser-Bessel
window \cite{kunis_potts2008,fourmont_2003} as the NFFT interpolating function.
Since the full precomputation of the window function at each node in
spatial and frequency domains requires lots of memory space, we have
used a tensor product form for the multivariate window function, that
requires only unidimensional precomputations. This method uses less
memory at the price of some extra multiplications
\cite{kunis_potts2008,fourmont_2003}. The accuracy
\cite{kunis_potts2008,fourmont_2003} of our simulation can be improved
 by increasing both the oversampling factor and the convolution
length, at the price of extra memory consumption and CPU time
(Table~\ref{Error_norms}).


\subsection{Simulation of lensed maps}

We applied our simulation algorithm of lensed CMB maps (both temperature
and polarization), as described in Section~\ref{sub:Algorithm},
to $1000$ independent realizations with HEALPix resolution $nside=1024$,
and a maximum multipole $l_{max}=2048$. In Figure~\ref{lensed_unlensed_CMB_map_nside1024},
we show one such realization of a lensed CMB temperature field, as
well as the difference between the lensed and unlensed fields. 



\begin{figure}[h]
\begin{centering}
\includegraphics[scale=0.25,angle=90]{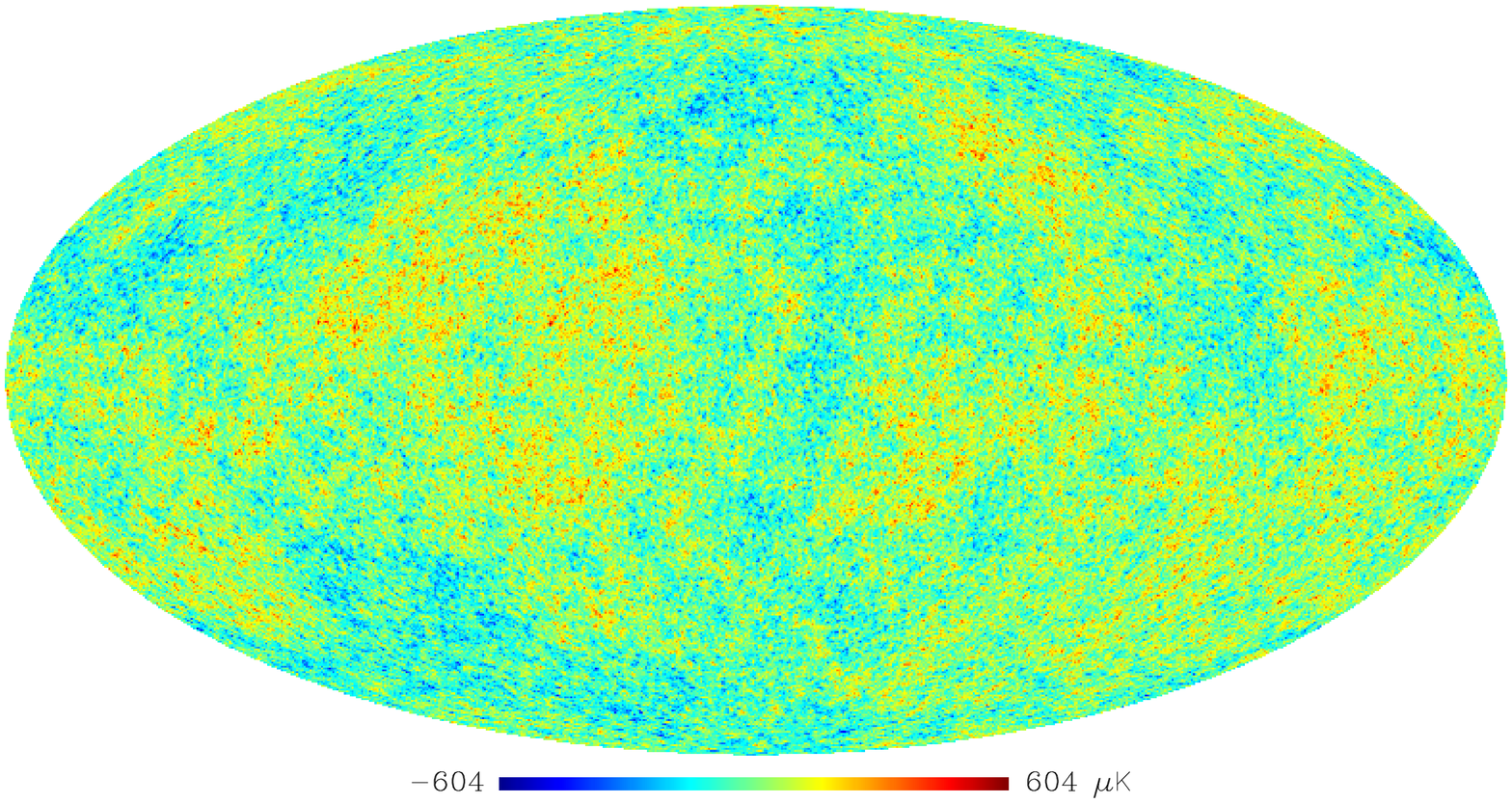}
\par\end{centering}
\begin{centering}
\includegraphics[scale=0.25,angle=90]{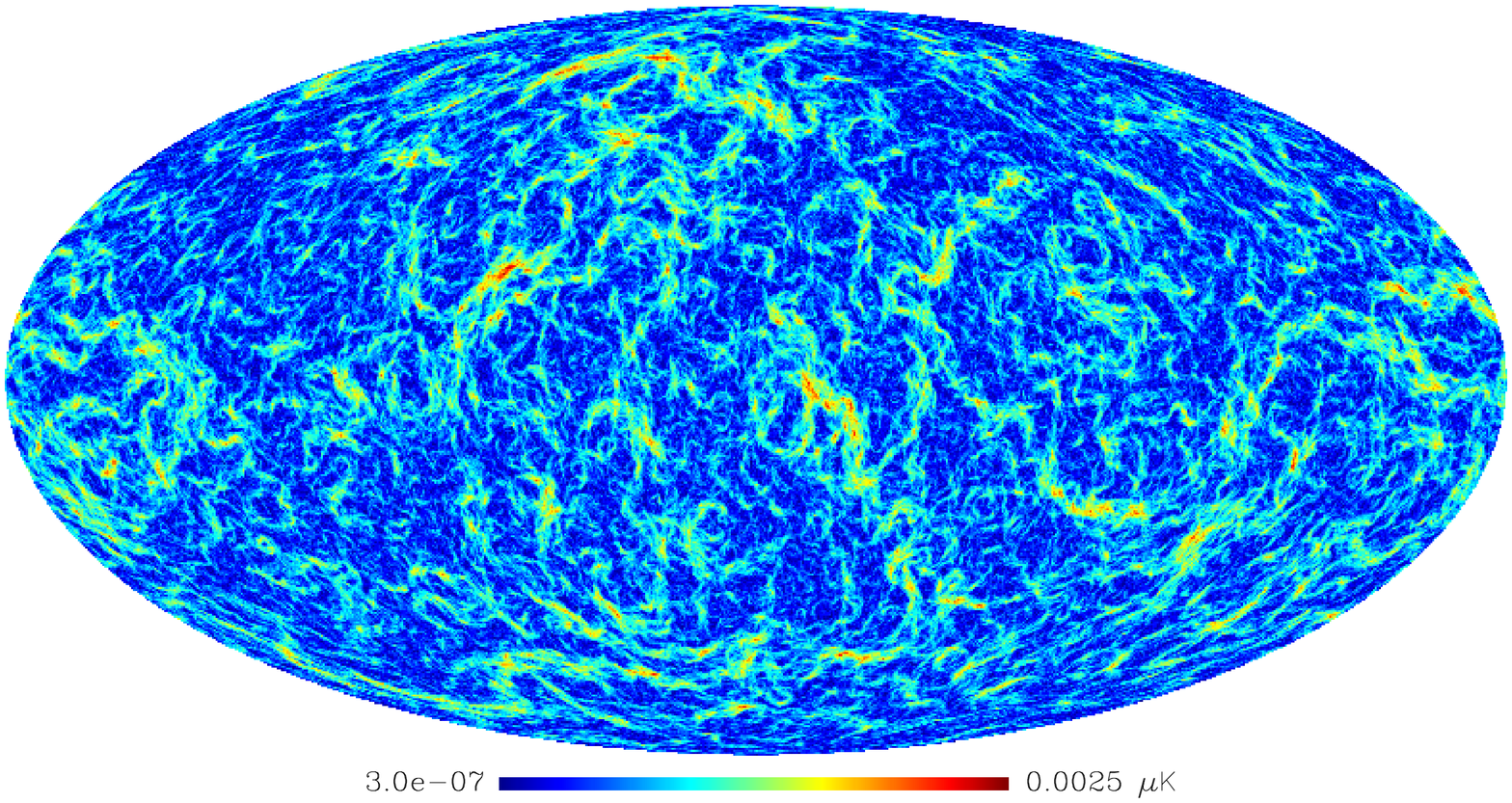}
\par\end{centering}
\begin{centering}
\includegraphics[scale=0.25,angle=90]{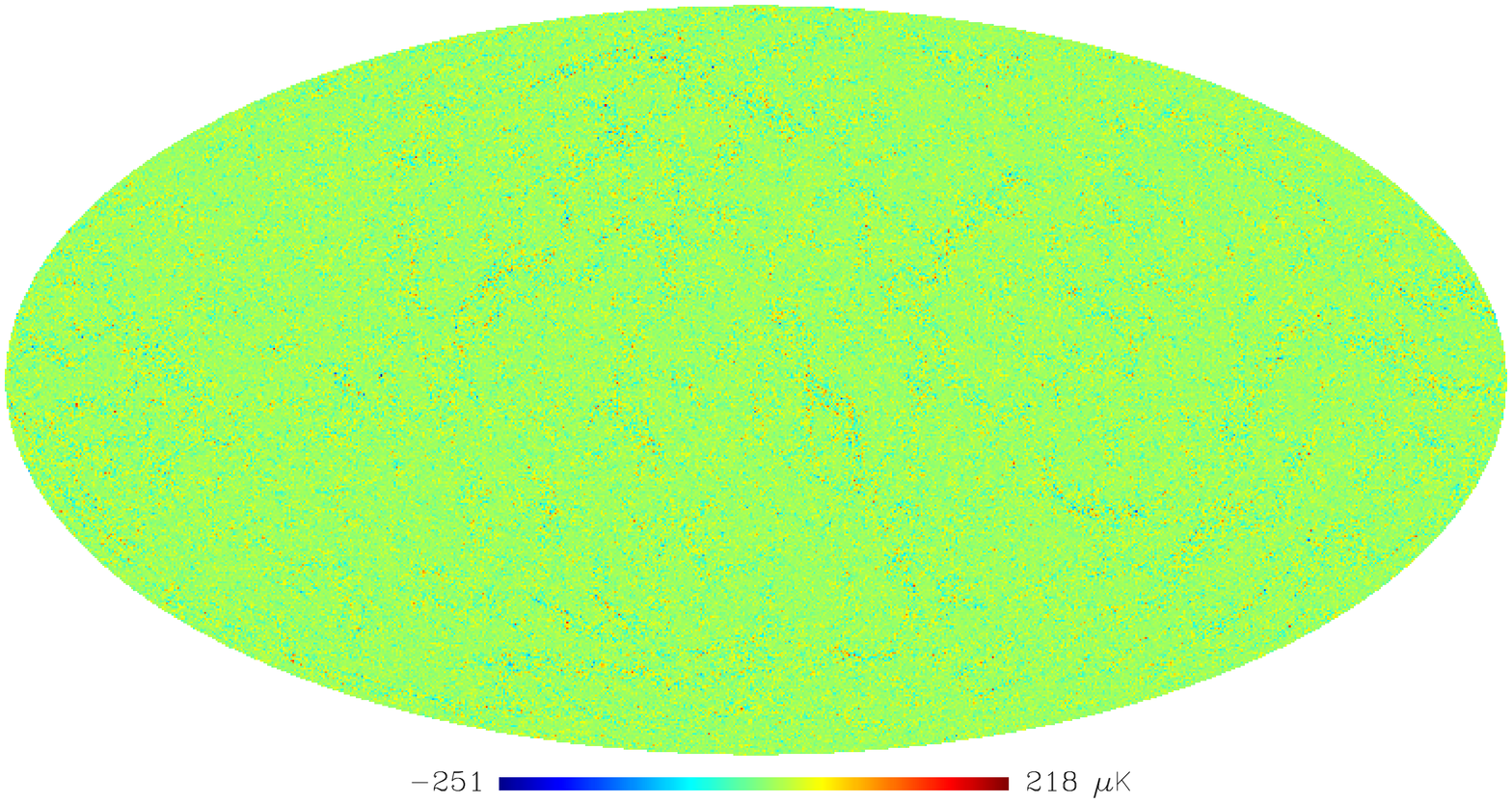}
\par\end{centering} \caption{(Top) A realization of lensed CMB map
(nside=1024). (Middle) A realization of the amplitude of the deflection
field (nside=1024). (Bottom) Difference of lensed and unlensed CMB maps
(nside=1024). These maps are obtained using NFFT for the oversampling
factor ($\sigma$)=2 and Convolution length $(K)=4$.}
\label{lensed_unlensed_CMB_map_nside1024}
\end{figure}

\begin{figure}[h]
\begin{centering}
\includegraphics[scale=0.25,angle=90]{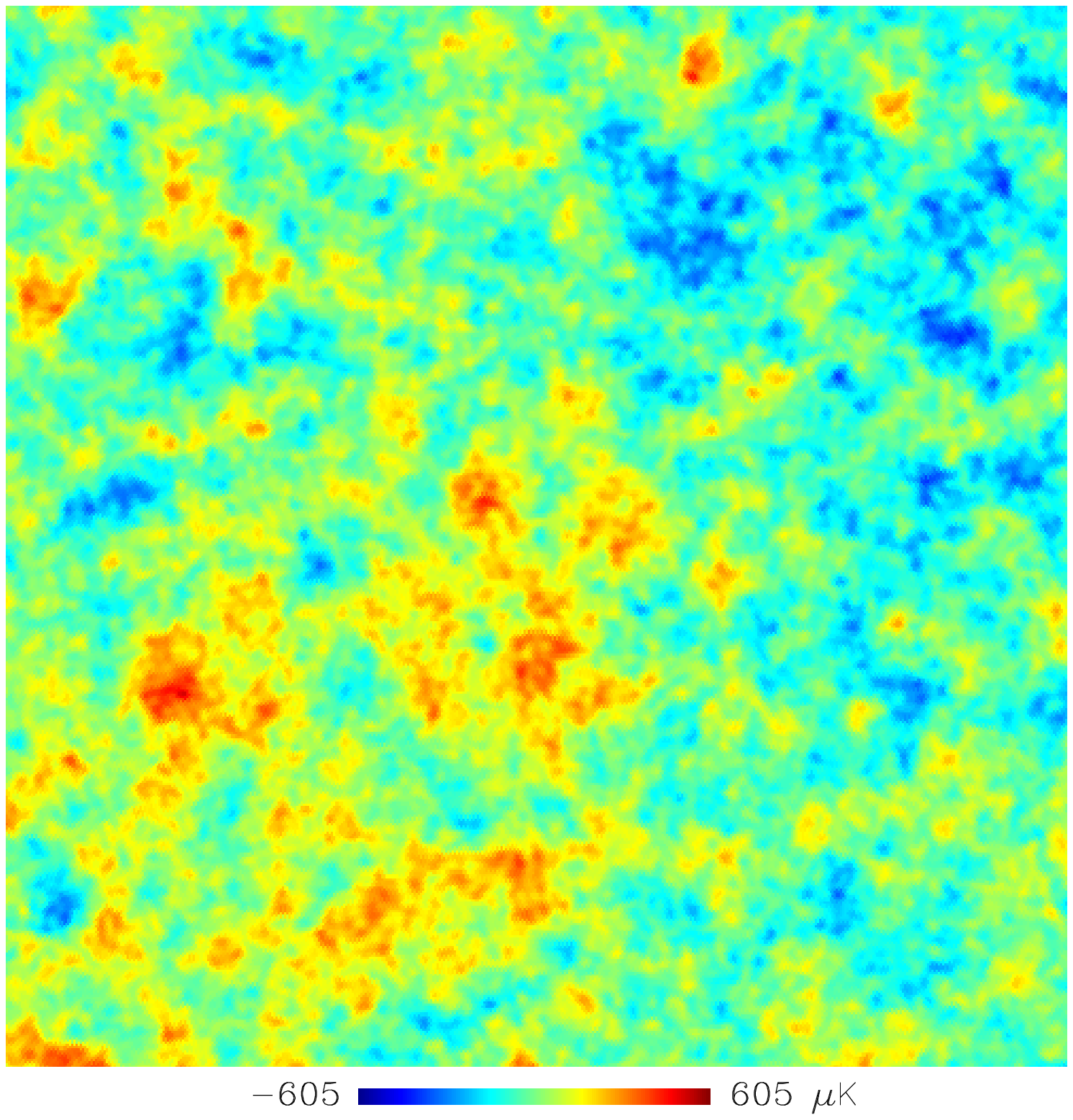}\hspace{0.1in}\includegraphics[scale=0.25,angle=90]{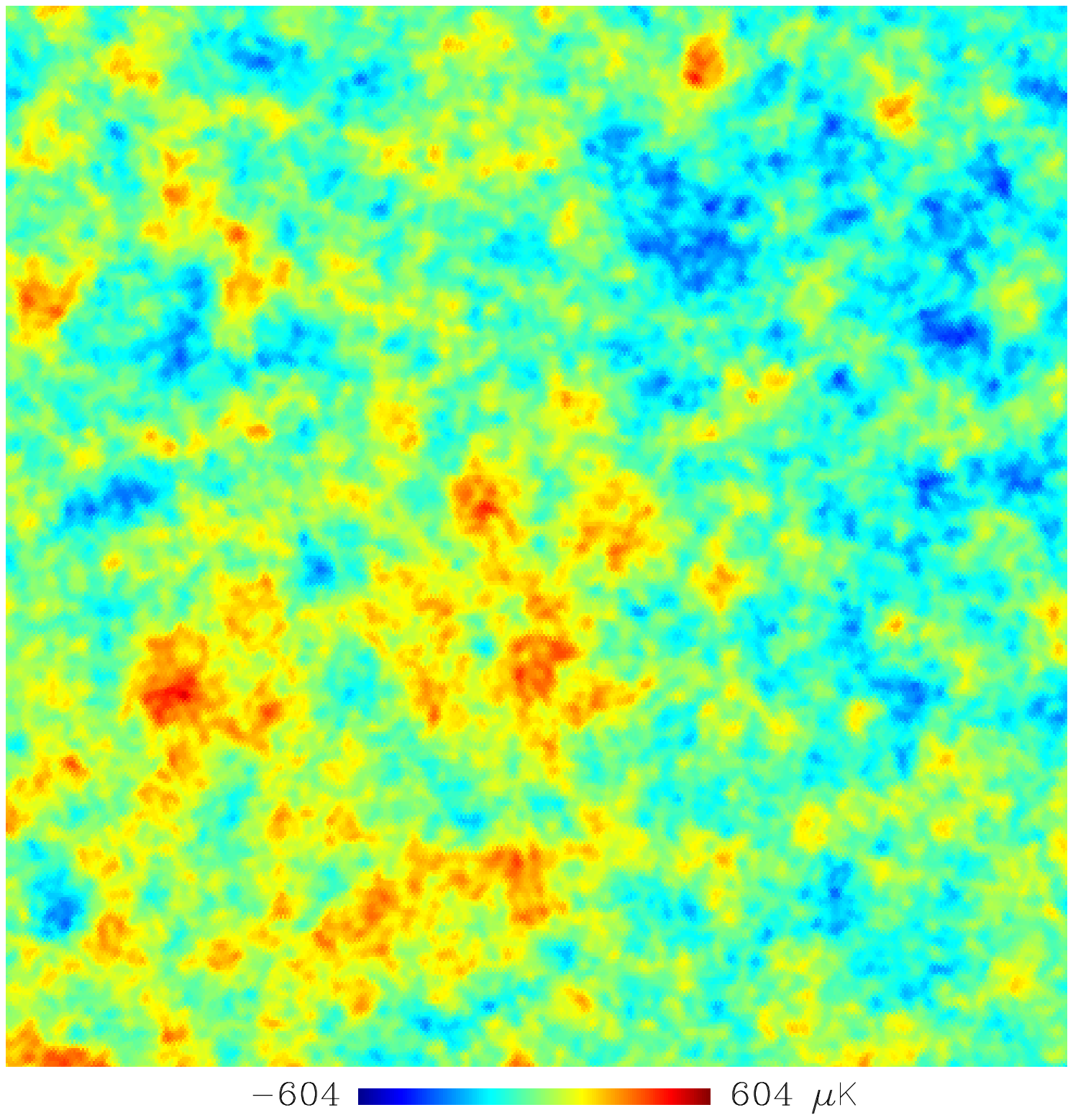}
\par\end{centering}

\begin{centering}
\includegraphics[scale=0.25,angle=90]{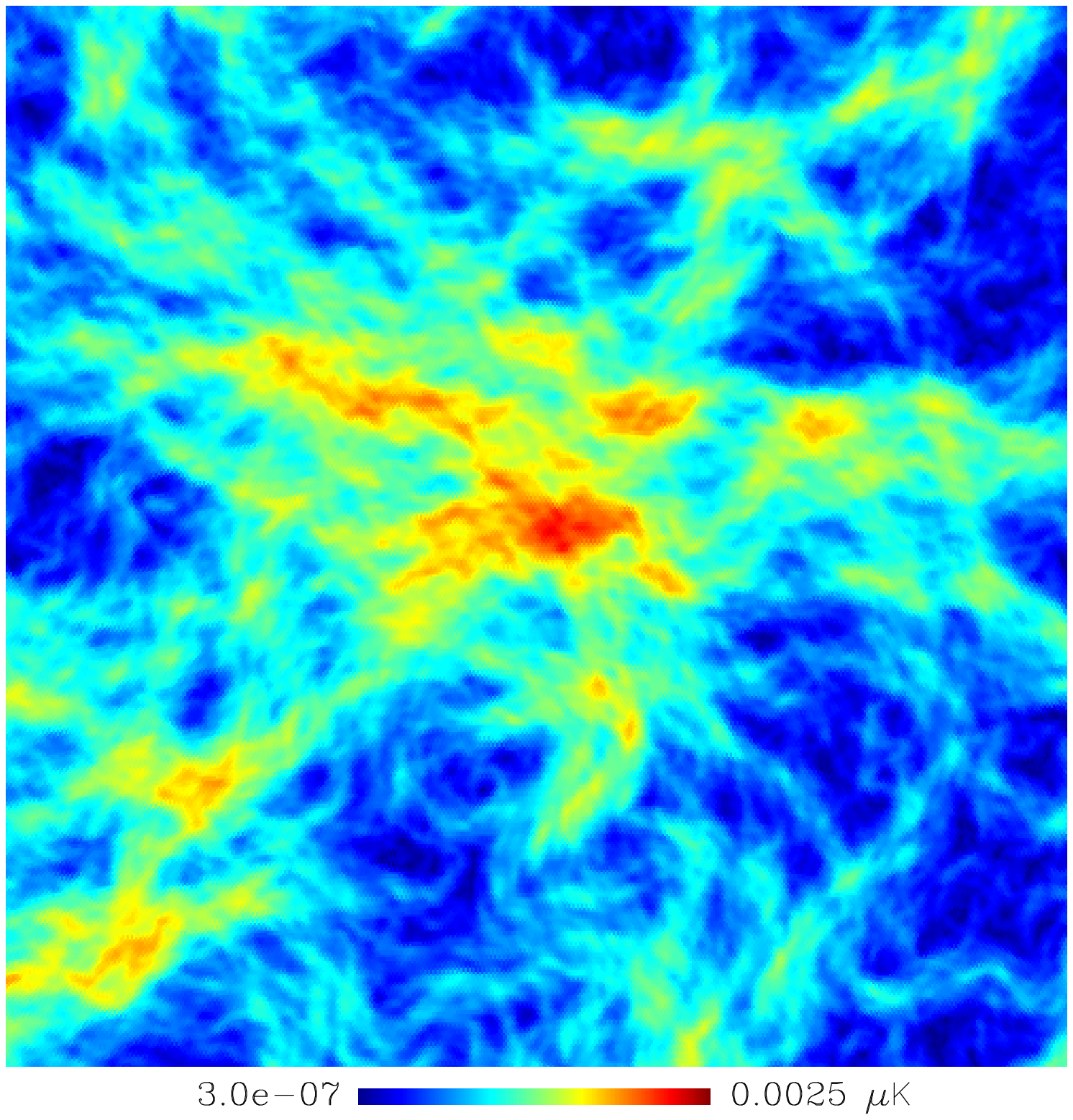}\hspace{0.1in}
\includegraphics[scale=0.25,angle=90]{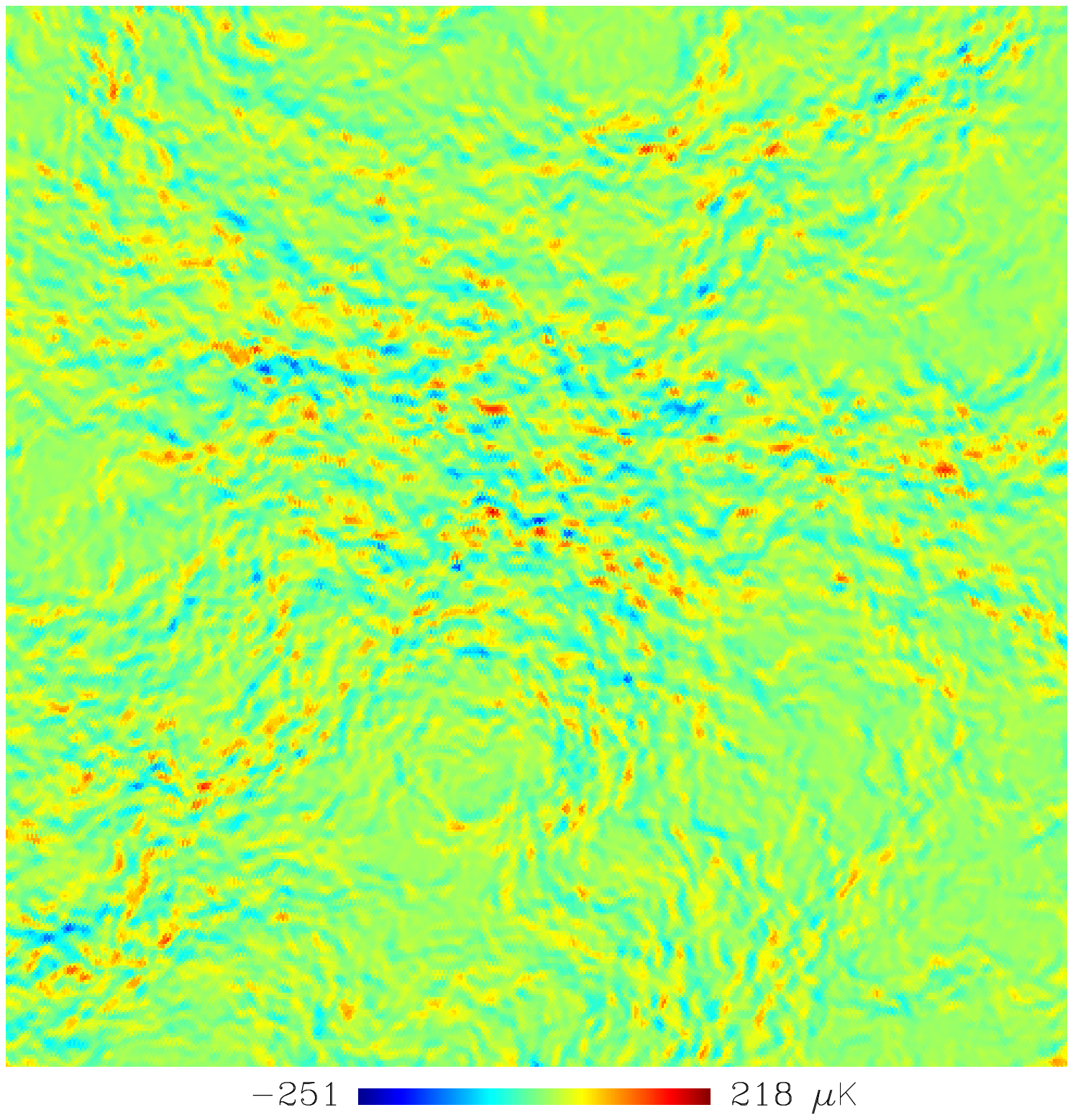}
\par\end{centering} \caption{(Top left) A small portion of a simulated
unlensed CMB temperature anisotropy map. (Top right). A small portion of
the corresponding lensed CMB temperature anisotropy map. (Bottom left) A small
portion of the amplitude of the simulated deflection field map. (Bottom
right) A small portion of the difference of simulated lensed and
unlensed CMB maps. These maps are obtained using NFFT for the
oversampling factor ($\sigma$)=2 and Convolution length $(K)=4$.}
\label{all_cutmap_nside1024.ps}
\end{figure}

Since weak lensing of CMB is a tiny effect at small angular scales, we
have shown a realization of a small portion of the unlensed CMB temperature
anisotropies, lensed CMB temperature anisotropies, amplitude of deflection
field and, the difference of lensed and unlensed CMB temperature
anisotropies in Figure \ref{all_cutmap_nside1024.ps} to better
illustrate the lensing effect. Although unlensed and lensed CMB temperature anisotropies
are indistinguishable to the naked eye, the correlation between the deflection
field and the difference of lensed and unlensed CMB temperature
anisotropies is clearly visible.

Table {[\ref{CPU_memory_with_res}] shows the typical CPU time and memory
that are required to simulate a single realization of unlensed and lensed
CMB temperature and polarization, with different resolutions.
Storage of the window function at the grid points both in spatial and
frequency domain before computing the Fourier transform consumes a fair
amount of memory, which ultimately increases the overall memory requirement
for the simulation of lensed CMB
maps\cite{kunis_potts2008,fourmont_2003}.

\begin{table}[h]
\caption{Variation of CPU time and memory requirements with resolution
 to simulate CMB maps (unlensed and lensed)[oversampling
 factor ($\sigma$)=2, Convolution length $(K)=4$] using NFFT.}
\begin{centering}
\begin{tabular}{|c|c|c|c|}
\hline 
nside&
$l_{max}$&
CPU&
Memory
\tabularnewline
&
&
time&
requirement
\tabularnewline
\hline 
256&
512&
1 min 12 sec&
491 MB
\tabularnewline
\hline 
512&
1024&
6 min 8 sec&
1.9 GB
\tabularnewline
\hline 
1024&
2048&
32 min &
7.6 GB
\tabularnewline
\hline
\end{tabular}
\label{CPU_memory_with_res} 
\par\end{centering}
\end{table}

Table {[\ref{CPU_memory_with_conv}] shows the same, but with different
convolution lengths. Increase of the convolution length not only increases
the computational cost of the interpolation part of NFFT, but also
increases the cost of the precomputation of window function and memory
requirement as one has to compute and store the window function at a
larger number of grid points in the spatial domain before doing
NFFT\cite{kunis_potts2008,fourmont_2003}.

\begin{table}[h]

\caption{Variation of CPU time and memory requirements with the
 convolution length $(K)$ for simulating a realization CMB map(unlensed
 and lensed)[nside$=1024$, $l_{max}=2048$] using NFFT.}
\begin{centering}
\begin{tabular}{|c|c|c|c|}
\hline 
Oversampling&
Convolution&
CPU&
Memory
\tabularnewline
factor&
length&
time&
requirement
\tabularnewline
$\left(\sigma\right)$&
$\left(K\right)$&
&

\tabularnewline
\hline 
2&
4&
32 min&
7.6 GB
\tabularnewline
\hline 
2&
6&
45 min&
8.4 GB
\tabularnewline
\hline 
2&
8&
60 min&
9.1 GB
\tabularnewline
\hline
\end{tabular}
 \label{CPU_memory_with_conv} 
\par\end{centering}
\end{table}

Figure~\ref{lensed_unlensed_cmb_cl} shows on the same plots the
theoretical power spectra $C_{l}^{XY}$, where $XY$ stands for
$TT,EE,TE,BB$ respectively, for the lensed and unlensed cases, as
predicted by CAMB. In the cosmological model we have chosen there
are no primordial tensors, hence $C_{l}^{BB}$ is entirely due to
lensing. %
\begin{figure*}
\begin{centering}
\includegraphics[scale=0.25]{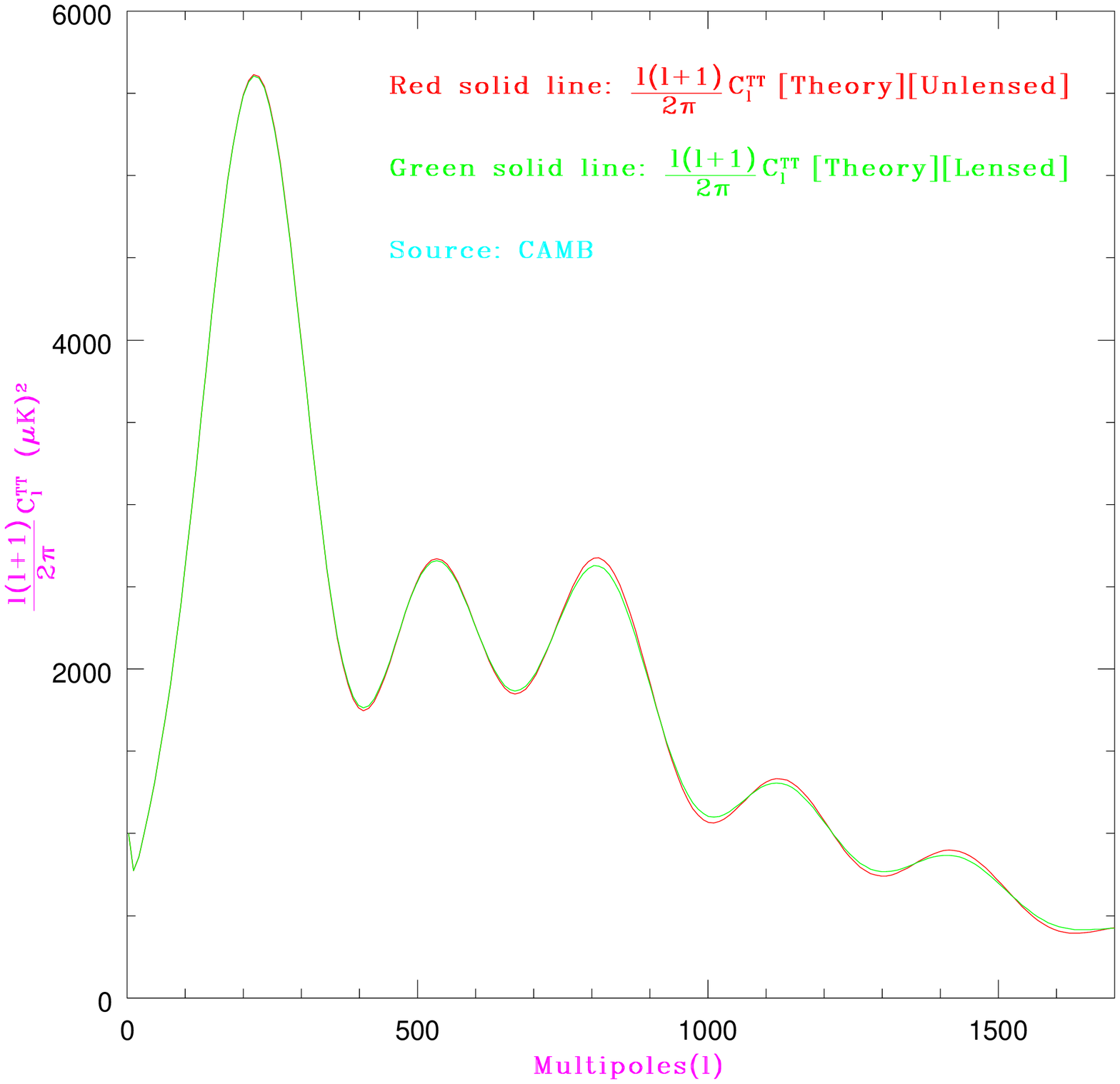}\includegraphics[scale=0.25]{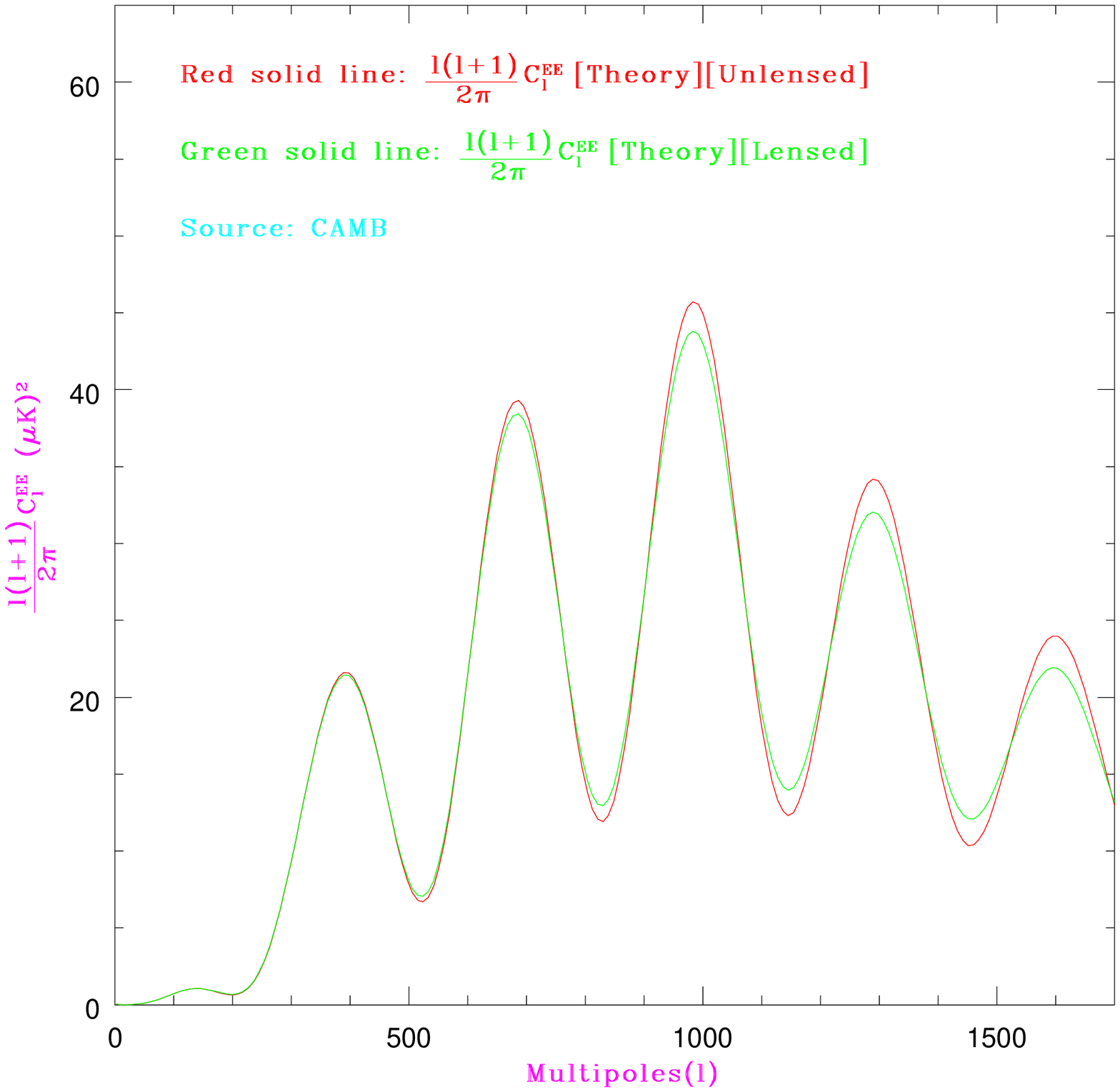}
\par\end{centering}

\begin{centering}
\includegraphics[scale=0.25]{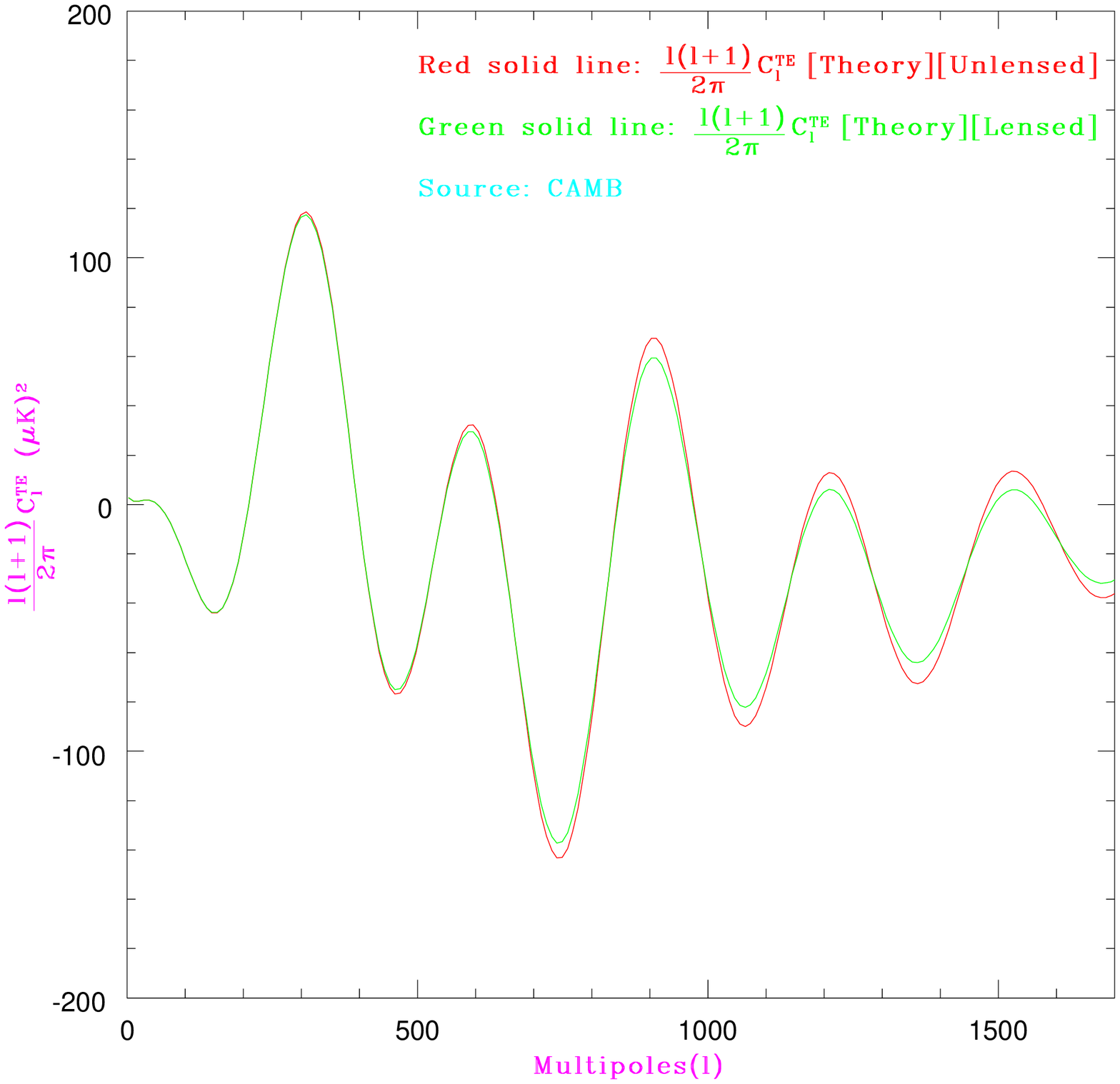}\includegraphics[scale=0.25]{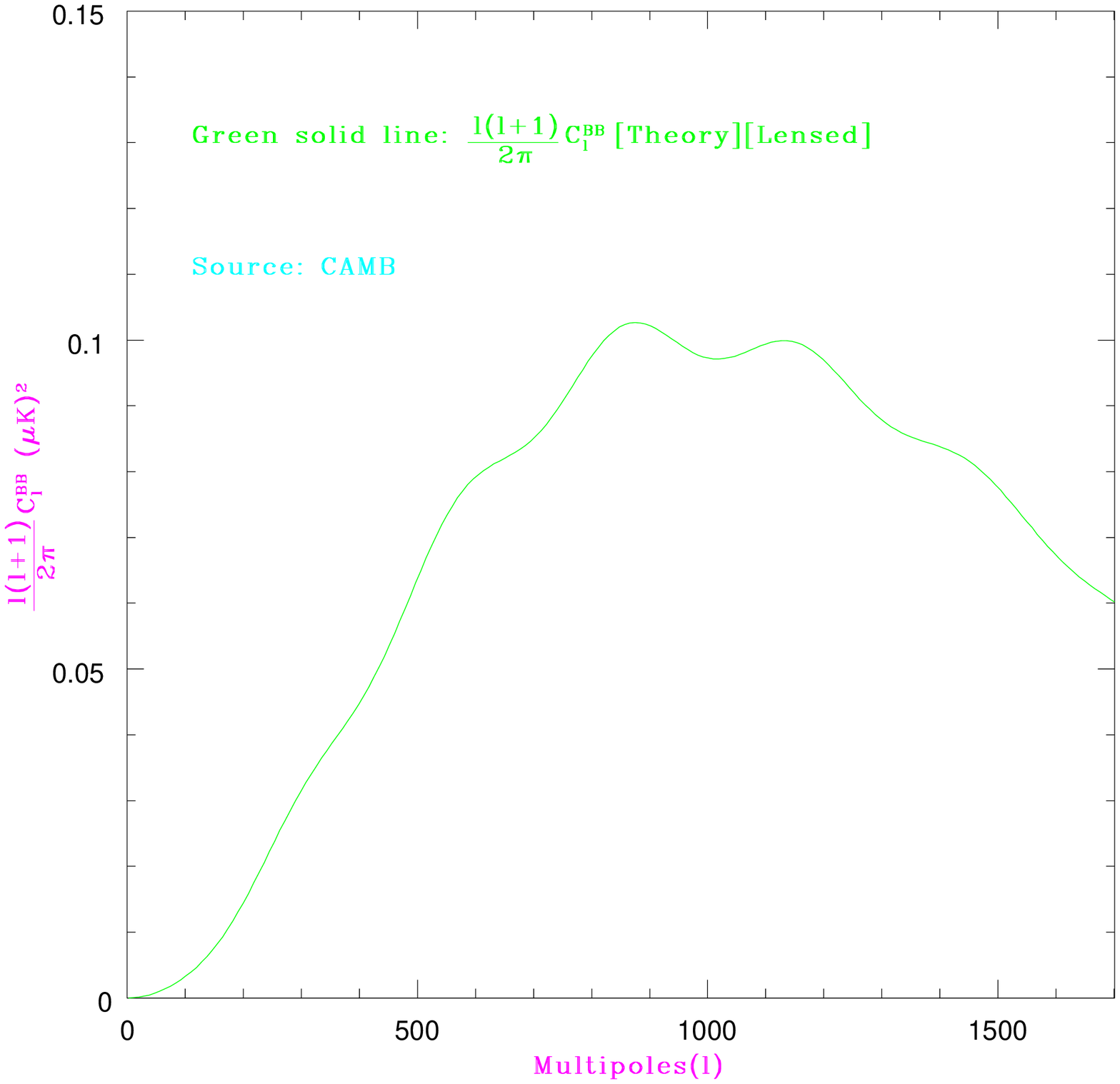}
\par\end{centering}

\caption{\textcolor{red}{Red} solid line is the theoretical angular power
spectrum of unlensed CMB, \textcolor{green}{Green} solid line is
the theoretical angular power spectrum of lensed CMB for temperature}

\label{lensed_unlensed_cmb_cl} 
\end{figure*}

An accurate recovery of this power spectrum from lensed polarization
maps is therefore a powerful test of our simulation method. In
Figure~\ref{lensed_cmb_cl} we show, on top of the lensed theoretical
spectra (solid lines), the average empirical power spectra computed from
the $1000$ simulations (dots). We can see that the agreement is
excellent, which is remarkable for $C_{l}^{BB}$ as explained above. We
have ignored the lensed angular power spectrum beyond the multipole
$l=1700$ in the comparison of average empirical power spectra and
theoretical power spectra because the accurate computation of the
average empirical power spectra for the multipoles $l > 1700$ requires
lensed CMB maps simulated from the power spectra of unlensed CMB and
lensing potential beyond the multipole $l=2048$.

\begin{figure*}

\begin{centering}
\includegraphics[scale=0.25]{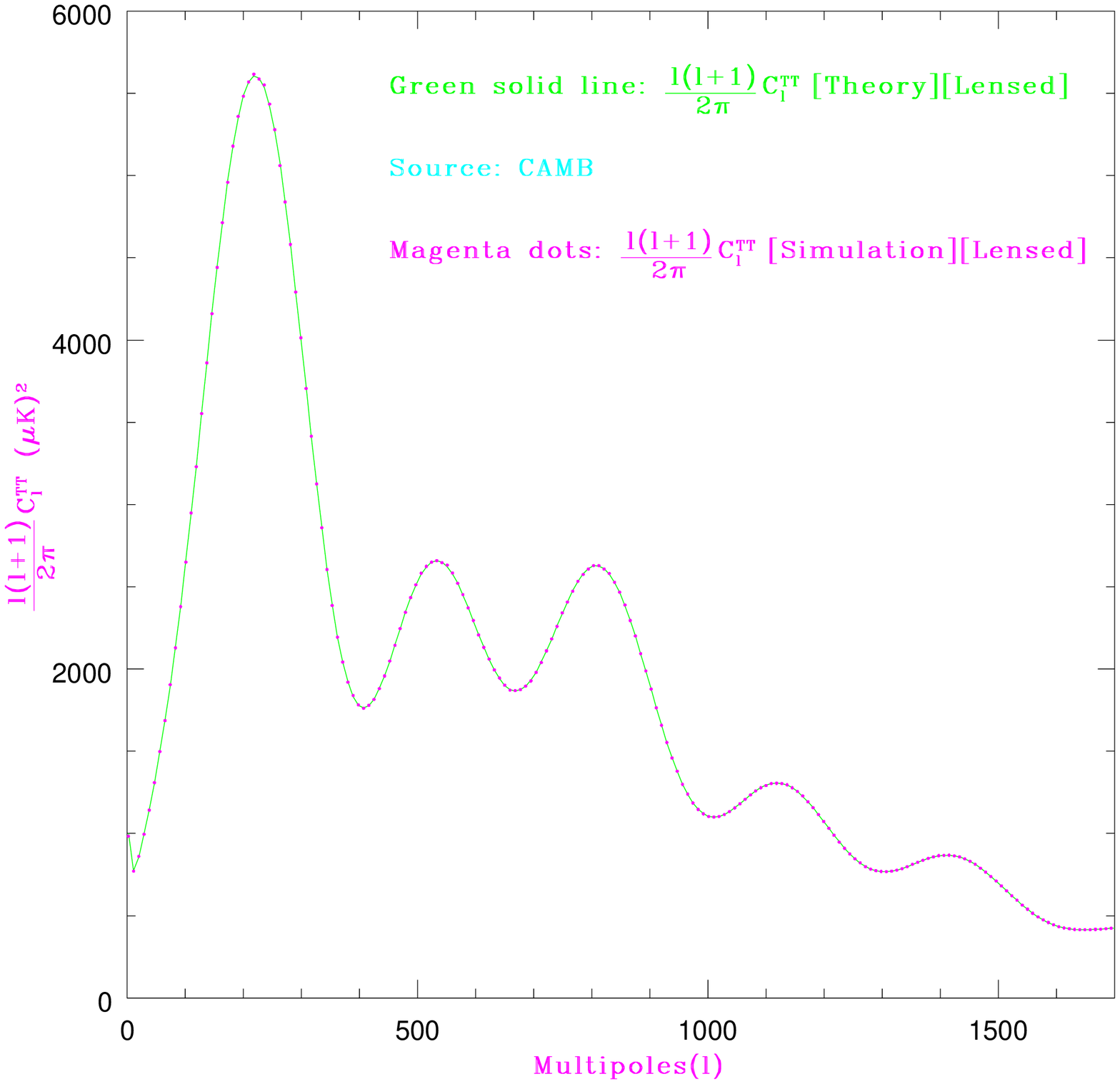}\includegraphics[scale=0.25]{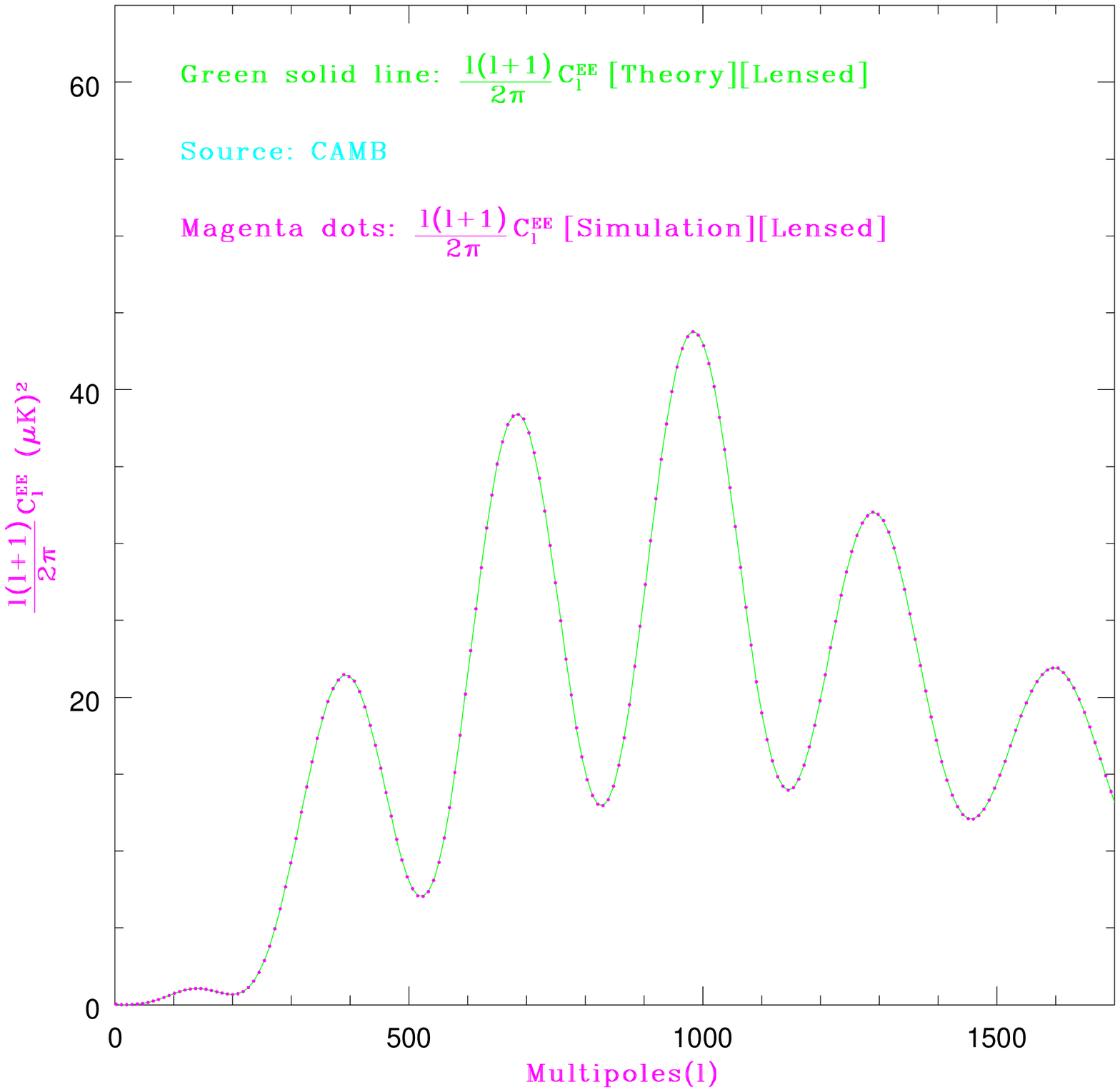}
\par\end{centering}

\begin{centering}
\includegraphics[scale=0.25]{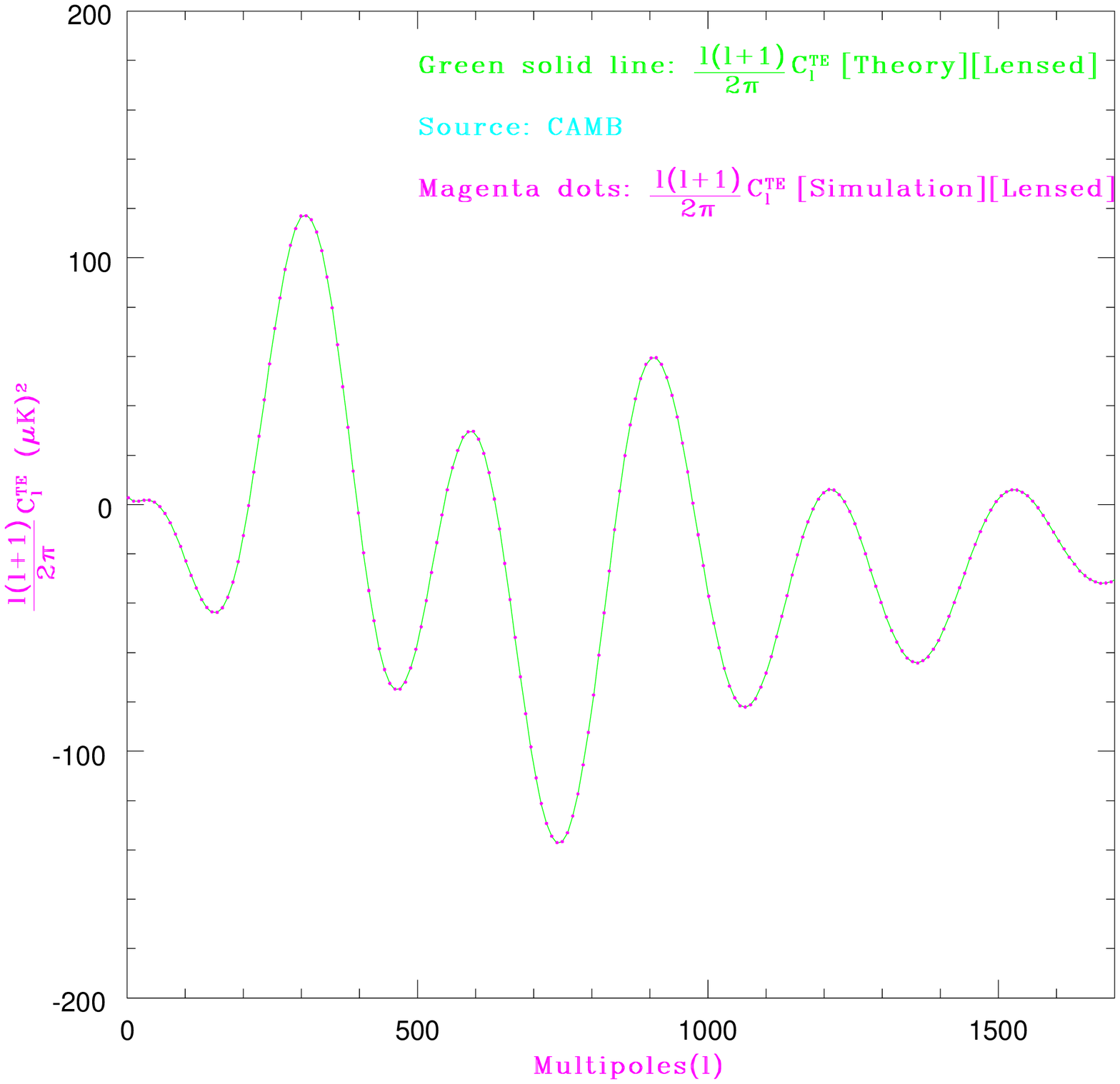}\includegraphics[scale=0.25]{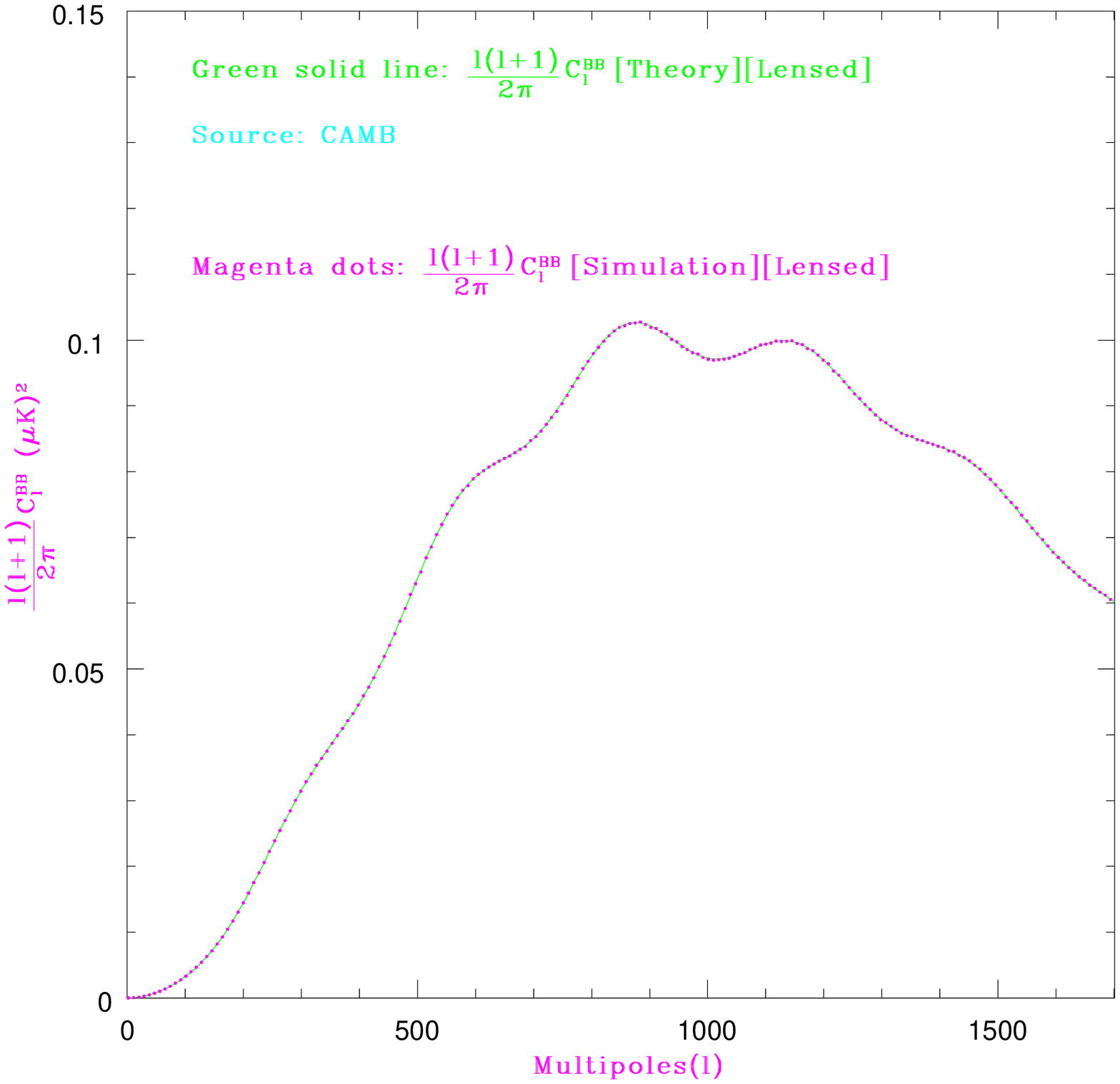}
\par\end{centering}

\caption{\textcolor{green}{Green} solid line is the theoretical angular
power spectrum $C_{l,\mathrm{th}}$ of lensed CMB, \textcolor{magenta}{Magenta}
dots are the average angular power spectrum $C_{l,\mathrm{simul}}$
recovered from 1000 realizations of lensed CMB maps (nside=$1024$
and $l_{max}=2048$)}

\label{lensed_cmb_cl} 
\end{figure*}

To have a more quantitative view of the accuracy of the method, we
show in Figure~\ref{diff_lensed_unlensed_cmb_cl} the relative difference
between the average empirical power spectra computed on the $1000$
simulations and the theoretical spectra from CAMB, both for the unlensed
(red) and lensed (green) cases. On each plot, we also show the theoretical
r.m.s. deviation of the averaged empirical spectra, computed neglecting
the small lensing-induced non-Gaussianity in the lensed cases. Note
that this corresponds to a very small underestimation of the scatter
(\cite{smith2006,rocher2008}). Taking into account the fact that
the averaged power spectra are nearly Gaussian distributed (due to
the central limit theorem), we can assess the presence of possible
biases in the recovered spectra by computing the reduced
$\chi^{2}$ statistics.

\begin{eqnarray}
Z^{2}_{XY}=\frac{N_{rlz}}{(l_{max}-1)}\sum^{l_{max}}_{l=2}\frac{(2 l +1)(C^{XY}_{l,\mathrm{simul}}-C^{XY}_{l,\mathrm{th}})^{2}}{\left[(C^{XY}_{l,\mathrm{th}})^{2}+C^{XX}_{l,\mathrm{th}}C^{YY}_{l,\mathrm{th}}\right]}\hspace{0.15in}
\end{eqnarray}

Here $N_{rlz}$ is the number of independent realizations of angular power
spectra under consideration.

\begin{table}[h]

\caption{Reduced $\chi^{2}$ statistics for the recovered unlensed angular
  power spectrum}
\begin{centering}
\begin{tabular}{|c|c|c|}
\hline 
Angular&
Value of&
\tabularnewline
power&
$\chi^{2}$&
$P(\infty>Z^{2}_{XY} \geq z^{2}_{XY})$
\tabularnewline
spectrum&
statistics&
\tabularnewline
$(C^{XY}_{l,\mathrm{simul}})$&
$(z^{2}_{XY})$&
\tabularnewline
\hline 
$C^{TT}_{l,\mathrm{simul}}$&
0.9574&
92\%
\tabularnewline
\hline 
$C^{EE}_{l,\mathrm{simul}}$&
0.9879&
65\%
\tabularnewline
\hline 
$C^{TE}_{l,\mathrm{simul}}$&
0.9901&
62\%
\tabularnewline
\hline
\end{tabular}
\label{chisquare_unlensed} 
\par\end{centering}
\end{table}
\begin{table}[h]
\caption{Reduced $\chi^{2}$ statistics for the recovered lensed angular
  power spectrum}
\begin{centering}
\begin{tabular}{|c|c|c|}
\hline 
Angular&
Value of&
\tabularnewline
power&
$\chi^{2}$&
$P(\infty>Z^{2}_{XY} \geq z^{2}_{XY})$
\tabularnewline
spectrum&
statistics&
\tabularnewline
$(C^{XY}_{l,\mathrm{simul}})$&
$(z^{2}_{XY})$&
\tabularnewline
\hline 
$C^{TT}_{l,\mathrm{simul}}$&
1.0030&
46\%
\tabularnewline
\hline 
$C^{EE}_{l,\mathrm{simul}}$&
0.9928&
58\%
\tabularnewline
\hline 
$C^{BB}_{l,\mathrm{simul}}$&
0.9928&
58\%
\tabularnewline
\hline
$C^{TE}_{l,\mathrm{simul}}$&
0.9933&
57\%
\tabularnewline
\hline
\end{tabular}
\label{chisquare_lensed} 
\par\end{centering}
\end{table}

Table~\ref{chisquare_unlensed} \& \ref{chisquare_lensed} show that the
probability of the reduced $\chi^{2}$ statistics $(Z^{2}_{XY})$ having
values greater than the estimated values $(z^{2}_{XY})$ are quite large
both for unlensed and lensed power spectrum. This strengthens our claim
about the unbiasedness in the simulation of unlensed and lensed CMB maps
using NFFT.

\begin{figure*}
\begin{centering}
\includegraphics[scale=0.25]{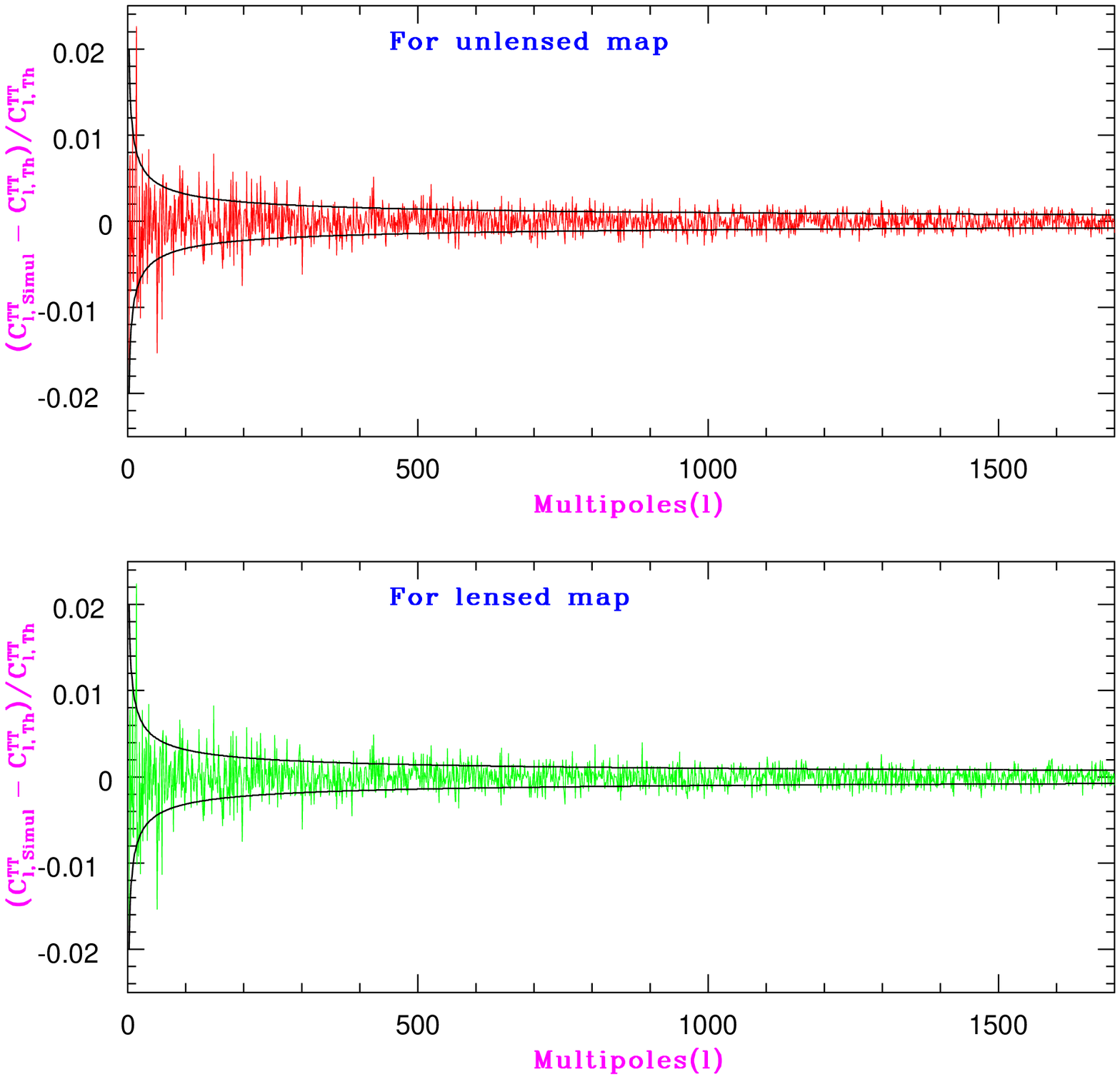}\includegraphics[scale=0.25]{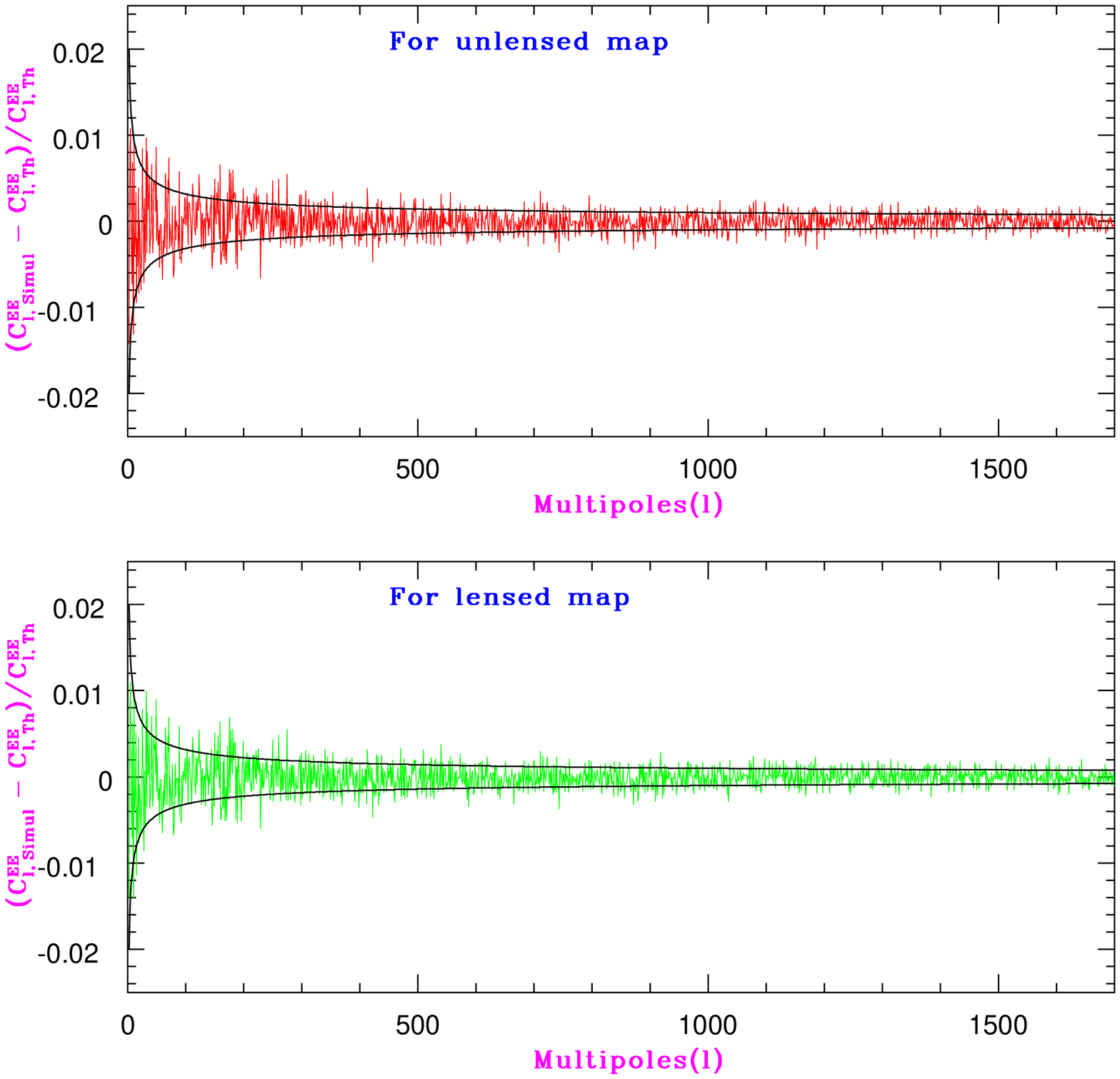}
\includegraphics[scale=0.25]{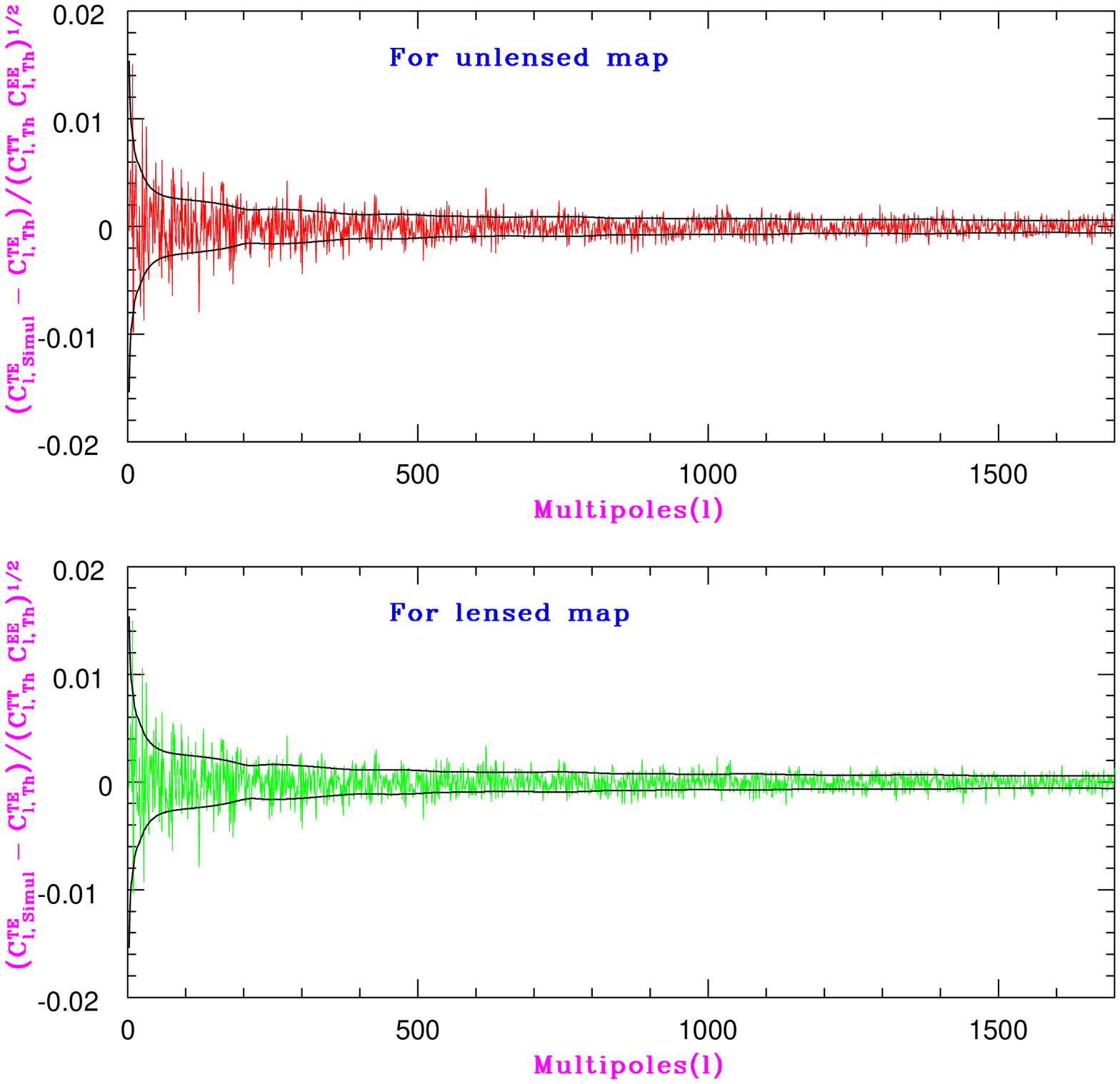}\includegraphics[scale=0.25]
{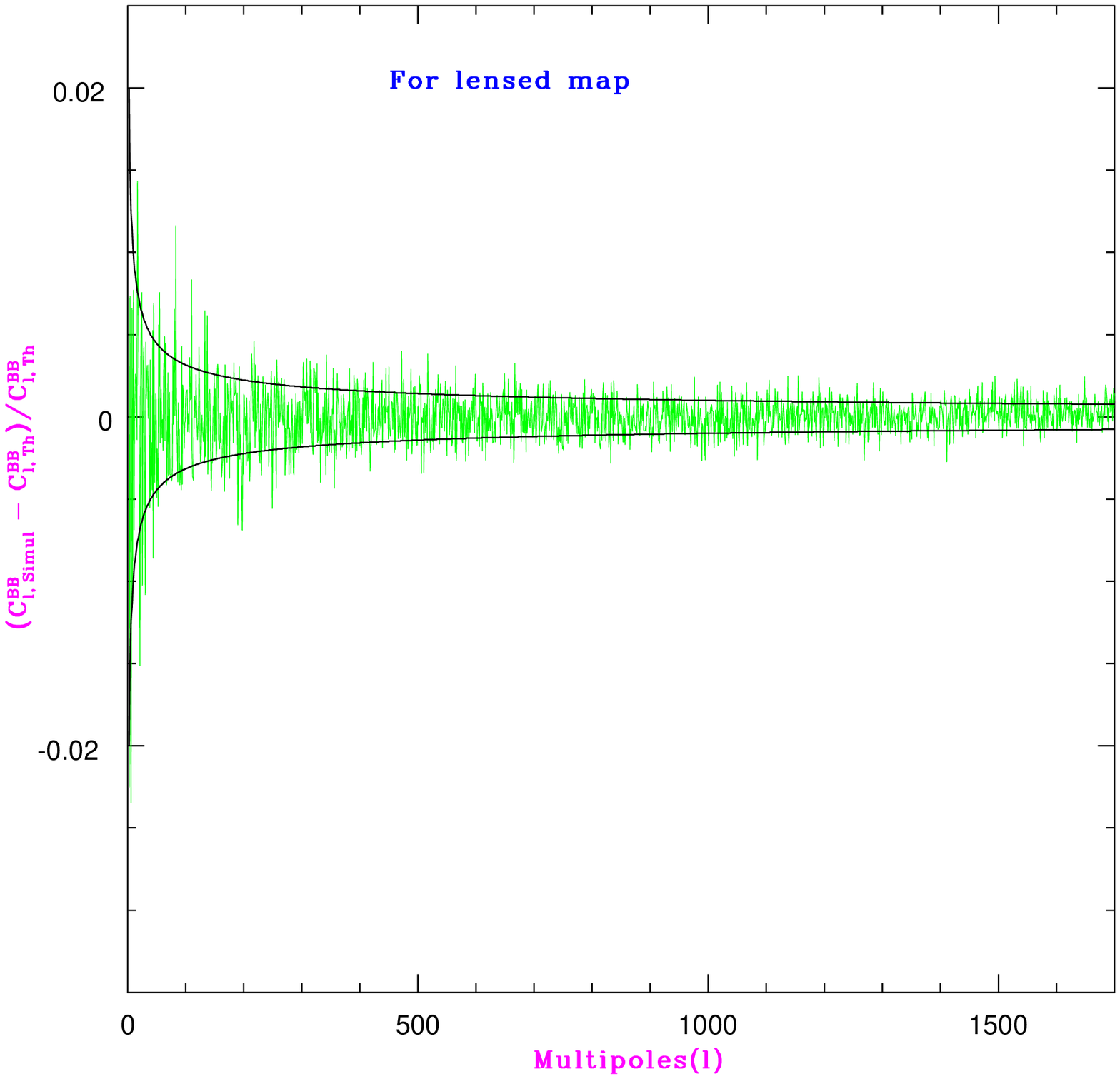} 
\par\end{centering}

\caption{Fractional difference of average angular power spectrua recovered
from $1000$ realizations of CMB maps (nside=$1024$ and $l_{max}=2048$)
and their corresponding theoretical angular power spectra.
\textcolor{red}{Red} lines are for unlensed maps and \textcolor{green}{green}
lines are for lensed maps. \textcolor{black}{Black} solid lines show
the theoretical cosmic variance.}

\label{diff_lensed_unlensed_cmb_cl} 
\end{figure*}

\section{Summary}
\label{summary} Accurate predictions for the expected CMB anisotropies
are required for analyzing future CMB data sets, which ultimately
require accurately simulated lensed maps. The most popular pixelization
used to analyze full-sky CMB maps is the HEALPix
pixelization. In order to simulate lensed CMB anisotropies at HEALPix
grid points we have to compute unlensed CMB anisotropies at irregularly
spaced grid points over the sphere, determined by the deflection field
and remapping equations.  Since remapping on a sphere can be recast into
remapping on a $2$-d torus, we have used the NFFT library to compute
lensed CMB anisotropies at HEALPix grid points and experimented with different
settings of the accuracy parameters. We have obtained that for a nside=$1024$ map
a $10^{-8}$ accuracy is easily reached when setting the $(\sigma,K)$ parameters to $(2,4)$. 
With our current implementation of the method, correspond to a $32$~min computation on 
a classical PC configuration. This can probably be improved by parallelizing the 
algorithm.
Furthermore, the average angular power spectra $C_{l,\,\mathrm{simul}}$  
recovered from 1000 realizations of lensed and unlensed CMB maps are
also found to be consistent with the corresponding theoretical ones,
$C_{l,\, th}$. This validates our simulation of lensed CMB
maps.

Such simulations will be a useful tool for the analysis and
interpretation of upcoming CMB experiments such as PLANCK and ACT.
However, they are not the only possible use of this technique. Indeed, the 
simulation of the lensing deflection field can be improved by going 
from the simple Born approximation to ray-tracing through dark matter 
N-Body simulations. Ray-tracing faces a similar problem as the simulation
of the lens effect on CMB maps, i.e. accurately resampling a vector field on the sphere.
Current state-of-the-art ray-tracing algorithms, like \cite{teyssier2008} could be 
made more accurate by using the technique described here.

\appendix

\section{Spin $s$ functions on a sphere and 2-d torus}

\label{spin_function} Spin $s$ square-integrable functions ${}_{s}f(\theta,\varphi)$
on a unit sphere are conveniently expanded in spin-weighted spherical
harmonics ${}_{s}Y_{lm}(\theta,\varphi)$ of same spin \cite{zaldarriaga1997,newman1996,goldberg1967}. \begin{eqnarray}
{}_{s}f(\theta,\varphi) & = & \sum_{l=0}^{l_{max}}\sum_{m=-l}^{l}{}_{s}f_{lm}\,\,{}_{s}Y_{lm}(\theta,\varphi)\label{spin_s_func_sphere}\end{eqnarray}
 with the inverse transform, \begin{eqnarray}
{}_{s}f_{lm} & = & \int_{\Omega}\, d\Omega\,\,{}_{s}f(\theta,\varphi)\,{}_{s}Y_{lm}^{*}(\theta,\varphi)\label{spin_s_flm_sphere}\end{eqnarray}

These harmonics, with $l\in\mathbb{N}$, $m\in\mathbb{Z}$ and max $\left(|m|,|s|\right)\le l$,
form an orthonormal basis for the decomposition of spin $s$ square-integrable
functions on the sphere. They are explicitly given in a factorized
form in terms of the Wigner rotation matrices $D_{m\, m'}^{l}(\varphi,\theta,\rho)$,
\begin{eqnarray}
{}_{s}Y_{lm}(\theta,\varphi) & = & (-1)^{s}\,\sqrt{\frac{2\, l+1}{4\pi}}\, D_{m\,(-s)}^{*\, l}(\varphi,\theta,0)\label{harm_slm}\end{eqnarray}

With our conventions for the Euler angles
\cite{edmonds1957,varshalovich1988}, we have, \begin{eqnarray} D_{m\,
m'}^{l}(\varphi,\theta,\rho) & = & e^{-i\, m\,\varphi}\,\, d_{m\,
m'}^{l}(\theta)\,\, e^{-i\, m'\,\rho}\label{wig_matrix}\end{eqnarray}

These rotation matrices (\ref{wig_matrix}) basically characterize
the rotation of spin-weighted spherical harmonics. Decomposition
shown in equation (\ref{wig_matrix}) is exploited by factoring the
rotation matrices in two separate rotation matrices as follows \cite{wiaux2006,mcewen2007},\begin{eqnarray}
D_{m\, m'}^{l}(\varphi,\theta,\rho)  =  \sum_{m''}\, D_{m\, m''}^{l}\left(\varphi-\frac{\pi}{2},-\frac{\pi}{2},\theta\right)\hspace{0.3in} \nonumber \\
 \hspace{0.5in} \times\,\,  D_{m''\, m'}^{l}\left(0,\frac{\pi}{2},\rho+\frac{\pi}{2}\right)\label{wig_decomp}\end{eqnarray}

Expressing the the Wigner rotation matrices (\ref{wig_matrix}) in
the above manner (\ref{wig_decomp}), equation (\ref{spin_s_func_sphere})
can be rewritten as, 
\begin{eqnarray}
{}_{s}f(\theta,\varphi)=\sum_{m=-l_{max}}^{l_{max}}\sum_{m'=-l_{max}}^{l_{max}}{}_{s}f_{m\,m'}\,e^{i(m\,\varphi+m'\,\theta)}
\label{spin_s_func_torus}
\end{eqnarray}
 where \begin{eqnarray}
{}_{s}f_{m\, m'} = \sum_{l=max(|m|,|m'|,|s|)}^{l_{max}}\,(-1)^{s}\,\sqrt{\frac{2\, l+1}{4\pi}}\,{}_{s}f_{lm}\hspace{0.4in}\nonumber \\
 \times \,\,\,\,d_{m'\, m}^{l}\left(\frac{\pi}{2}\right)\, d_{m'\,
  (-s)}^{l}\left(\frac{\pi}{2}\right)\,\exp\left[-i(m+s)\frac{\pi}{2}\right]\hspace{0.2in}
\label{spin_s_fmn_torus}\end{eqnarray}

The advantage of factoring the rotation matrices in this manner is
that now the Euler angles only occur in complex exponentials and we
need to evaluate $d_{m\, m'}(\theta)$ at $\theta=\frac{\pi}{2}$
only \cite{mcewen2007,edmonds1957,varshalovich1988,risbo1996,challinor2000}.

Computation of ${}_{s}f(\theta,\varphi)$ using equation
(\ref{spin_s_func_torus}) may not be the most efficient way, but the
presence of exponentials may be exploited such that techniques of fast
Fourier transform either on irregular or regular grid may be used for
rapid computation of double summations simultaneously. In both cases,
the domain of spin $s$ function ${}_{s}f(\theta,\varphi)$ must be
extended from the sphere, $(\theta,\varphi)\in[0,\pi]\times[0,2\pi]$ to
the 2-dimensional torus, $(\theta,\varphi)\in[0,2\pi]\times[0,2\pi]$
using the symmetry ${}_{s}Y_{lm}(2\pi-\theta,\pi+\varphi)=(-1)^{s}\,
{}_{s}Y_{lm}(\theta,\varphi)$ of spin-weighted spherical harmonics so that
equation (\ref{spin_s_func_torus}) becomes a complex to complex Fourier
transform over 2-dimensional torus. The computation of ${}_{s}f_{m\,
m'}$ for $\left(|m|,|m'|\right)\le l_{max}$, involves performing a
1-dimensional summation over a 2-dimensional grid, hence it is of order
${\cal O}(l_{max}^{3})$.

\section{CMB fields on 2-d torus}
\label{cmb_torus}
Factoring the rotation matrices in two separate rotation matrices
(\ref{wig_matrix}) and extending the domain of CMB, lensing potential
and defection fields from sphere to 2-d torus, equations (\ref{temp_sphere},\ref{pol_sphere},\ref{pot_sphere},\ref{deflec_sphere})
can be rewritten as,

\begin{eqnarray}
T(\theta,\varphi) =
 \sum_{m=-l_{max}}^{l_{max}}\sum_{m'=-l_{max}}^{l_{max}}T_{m\,
 m'}\,\,e^{i(m\,\varphi+m'\,\theta)}
\label{temp_torus}
\end{eqnarray}

\begin{eqnarray}
P(\theta,\varphi) =
 \sum_{m=-l_{max}}^{l_{max}}\sum_{m'=-l_{max}}^{l_{max}}P_{m\,m'}\,\,e^{i(m\,\varphi+m'\,\theta)}
\label{pol_torus}
\end{eqnarray}
\begin{eqnarray}
\Phi(\theta,\varphi) =
 \sum_{m=-l_{max}}^{l_{max}}\sum_{m'=-l_{max}}^{l_{max}}\Phi_{m\,
 m'}\,\,e^{i(m\,\varphi+m'\,\theta)}
\label{pot_torus}
\end{eqnarray}
\begin{eqnarray}
\left(d_{\theta}+\, i\, d_{\varphi}\right)\,(\theta,\varphi)\hspace{1.8in}\nonumber \\ =\sum_{m=-l_{max}}^{l_{max}}\sum_{m'=-l_{max}}^{l_{max}}G_{m\,m'}\,\,e^{i(m\,\varphi+m'\,\theta)}\label{defl_torus}
\end{eqnarray}
and the corresponding Fourier modes are given by,
\vspace{0.1in}
\begin{eqnarray}
T_{m\, m'} = \sum_{l=max(|m|,|m'|)}^{l_{max}}\sqrt{\frac{2\, l+1}{4\pi}}\,\, T_{lm}\hspace{1.25in}\nonumber \\
  \times \,\,d_{m'\, m}^{l}\left(\frac{\pi}{2}\right)\,
   d_{m'\,0}^{l}\left(\frac{\pi}{2}\right)\,\exp\left[-im\frac{\pi}{2}\right]\hspace{0.25in}
\label{temp_harm_torus}
\end{eqnarray}

\vspace{0.1in}
\begin{eqnarray}
P_{m\, m'}  =  \sum_{l=max(|m|,|m'|,2)}^{l_{max}}\,\sqrt{\frac{2\, l+1}{4\pi}}\,(E_{lm}+i\,B_{lm})\hspace{0.5in}\nonumber\\
  \times  \,\,d_{m'\, m}^{l}\left(\frac{\pi}{2}\right)\,
   d_{m'\,(-2)}^{l}\left(\frac{\pi}{2}\right)\,\exp\left[-im\frac{\pi}{2}\right]\hspace{0.25in}
\label{pol_harm_torus}
\end{eqnarray}

\vspace{0.1in}
\begin{eqnarray}
\Phi_{m\, m'}  =  \sum_{l=max(|m|,|m'|)}^{l_{max}}\sqrt{\frac{2\, l+1}{4\pi}}\,\Phi_{lm}\hspace{1.25in}\nonumber \\
  \times \,\, d_{m'\,
   m}^{l}\left(\frac{\pi}{2}\right)\,d_{m'\,0}^{l}\left(\frac{\pi}{2}\right)\,\exp\left[-im\frac{\pi}{2}\right] 
\hspace{0.25in}
\label{pot_harm_torus}
\end{eqnarray}

\vspace{0.1in}
\begin{eqnarray}
G_{m\, m'}  =  \sum_{l=max(|m|,|m'|,1)}^{l_{max}}(-i)\sqrt{\frac{l\,(l+1)\,(2\, l+1)}{8\pi}}\,\Phi_{lm}\hspace{0.35in}\nonumber \\
  \times \,\,d_{m'\, m}^{l}\left(\frac{\pi}{2}\right)\,
   d_{m'\,(-1)}^{l}\left(\frac{\pi}{2}\right)\,\exp\left[-im\frac{\pi}{2}\right]\hspace{0.25in}
\label{field_harm_torus}
\end{eqnarray}

\vspace{0.15in}
\begin{acknowledgements}
We acknowledge the use of the HEALPix package for our map
pixelization. We also acknowledge the use of the NFFT package for our
work. We thank Eric Hivon for helpful discussions and
suggestions. One(SB) of the author's research at the \textit{Institut
d'Astrophysique de Paris} was supported by the \textit{Indo-French
centre for promotion of advanced scientific research} (CEFIPRA) through
grant 3504-3. SB thanks Francois R. Bouchet and Tarun Souradeep for their
constant encouragement and support throughout. 

\end{acknowledgements}

\end{document}